# Content Moderation Futures

Lindsay Blackwell

## 1.1 Abstract

This study examines the failures and possibilities of contemporary social media governance through the lived experiences of various content moderation professionals. Drawing on participatory design workshops with 33 practitioners in both the technology industry and broader civil society, this research identifies significant structural misalignments between corporate incentives and public interests. While experts agree that successful content moderation is principled, consistent, contextual, proactive, transparent, and accountable, current technology companies fail to achieve these goals, due in part to exploitative labor practices, chronic underinvestment in user safety, and pressures of global scale. I argue that successful governance is undermined by the pursuit of technological novelty and rapid growth, resulting in platforms that necessarily prioritize innovation and expansion over public trust and safety. To counter this dynamic, I revisit the computational history of care work, to motivate present-day solidarity amongst platform governance workers and inspire systemic change.

## 1.2 Introduction

Online platforms are essential venues for social interaction. As social media platforms[1] grow in size, reach, and influence, the question of how to govern human behavior at scale becomes increasingly critical. Contemporary platform governance hinges on what Roberts (2016) deems *commercial content moderation*, a complex assemblage of human labor, automated technologies, and organizational practices intended to ensure user safety through the scaled enforcement of standardized rules. Despite their growing prominence, scaled content moderation systems are frequently characterized by opacity, inconsistency, and ineffectiveness, particularly for users whose identities, histories, and experiences are not represented by dominant ideological structures.

---

[1] Tarleton Gillespie (2018) defines social media platforms as "online sites and services that (a) host, organize, and circulate users' shared content or social interactions for them, (b) without having produced or commissioned the bulk of the content, (c) built on an infrastructure beneath that circulation of information, for processing data for customer service, advertising, and profit."

The rise of social media platforms has led to unprecedented challenges in moderating human behavior at scale, with billions of users generating posts, images, and videos daily. Platforms like Facebook, YouTube, TikTok, and Reddit face mounting pressure from governments, advertisers, and users to prevent harm while also protecting expression. Though numerous scholars have examined the sociopolitical implications of content moderation—from the algorithmic biases embedded in automated enforcement tools (Noble, 2018) to the material conditions of outsourced moderators (Roberts, 2016; Roberts, 2019)—comparatively little research has engaged directly with the broad spectrum of professionals responsible for implementing and interrogating scaled platform governance. While content moderation professionals—including both industry and civil society workers, spanning roles across policy, operations, research, engineering, and other relevant domains—possess deep, practical knowledge of the governance systems they help enact, their experiences remain largely absent from both public and scholarly discourse.

Through a series of participatory design workshops with 33 content moderation professionals, this research surfaces a situated understanding of the values, challenges, and contradictions that shape contemporary social media governance. While experts agree that successful content moderation is principled, consistent, contextual, proactive, transparent, and accountable, technology companies repeatedly fail to achieve these goals at scale. The impractical pursuit of universal solutions to global governance yields vague, Western-centric policies, marginalizing non-dominant perspectives in ways further compounded by the technical limitations of automated enforcement systems. Business incentives conflict with user safety goals, resulting in chronic underinvestment and exploitative labor practices—and leaving individual workers feeling overburdened and isolated from their closest peers. I argue that successful governance is undermined by the pursuit of technological novelty and rapid growth, resulting in platforms that necessarily prioritize innovation and expansion over public trust and safety. I conclude by revisiting the computational history of care work, positioning worker solidarity as the foundation for meaningful structural change.

## 1.3 Related work

### 1.3.1 Failures of contemporary social media governance

Platforms rely on a combination of human labor and automated tools to enforce policies and guidelines, adhere to regulatory compliance obligations, and manage reputational risk (Gillespie, 2018; Roberts, 2019; Tyler et al., 2025). Social media companies typically publish formal policies for appropriate use, such as Meta's Community Standards[2], which describe specific behaviors that are and are not allowed on Facebook, Instagram, Messenger, and Threads. Users whose behavior is determined to violate these policies are subject to a variety of sanctions, including content removal and temporary or permanent account suspension.

Enforcing these policies at a global scale has proven to be an impossible challenge (Gillespie, 2018; Roberts, 2019). Many social media users are not aware that rules exist; a user is often first exposed to platform rules when they have violated one, often inadvertently (Chandrasekharan et al., 2018; Katsaros et al., 2022; Tyler et al., 2025). Other users may have an ambient awareness of the existence of site policies, but few can articulate specific rules or examples of inappropriate behavior, a problem intensified by the lack of transparency social media companies provide into their specific enforcement practices (Gillespie, 2018; Suzor, 2019). Rules also vary widely across platforms and often invoke complex or specialized terminology (Pater et al., 2016; Jiang et al., 2020), further exacerbating the burden on individual users to sufficiently understand platform rules and any consequences of breaking them.

These private rules—which regulate users' speech, conduct, and access—are enforced through scaled content moderation practices, or the systems and processes used to monitor and evaluate large volumes of user-generated text, images, and videos (Gillespie, 2018; Roberts, 2019). Contemporary platform governance ecosystems rely heavily on *commercial content moderation* (Roberts, 2016), or the process of outsourcing scaled content moderation labor to external service providers in lower-wage regions of the Majority World. A typical commercial content moderator reviews hundreds of pieces of content per day, earning low wages despite unrealistic performance standards and constant exposure to objectionable content (Roberts, 2016; Roberts, 2019). Moderators are expected to make individual enforcement decisions in a matter of seconds, requiring memorization of complex policies that frequently change.

---

[2] http://transparency.meta.com/policies/community-standards/

While users can report content directly to platforms for potential review (Crawford & Gillespie, 2016; Tyler et al., 2025), many technology companies leverage machine learning algorithms to automatically detect—and in some cases, enforce against—potentially violative content (West, 2018; Gorwa et al., 2020; Vaccaro et al., 2020). Developing such models requires a similarly invisible workforce to generate sufficiently large volumes of labeled training data (Gray & Suri, 2019), with platforms like MTurk facilitating the on-demand employment of data labeling workers, whose labor is required to translate complex cultural context into a format "legible" to computers (Irani, 2023).

**1.3.2 Promoting social justice under platform capitalism**

The purported promises of scaled content moderation have been increasingly undermined as social media platforms repeatedly fail to adequately prevent harm, both to individuals and society at large. Interdisciplinary scholarship has highlighted numerous ways in which scaled moderation systems fail, both due to insufficient technical systems and the structural, economic, and epistemological assumptions embedded within them (Gillespie, 2018; West et al., 2019; Gorwa et al., 2020). Despite increasing public pressure to discourage and even limit discriminatory or otherwise threatening speech (Grimmelmann, 2013), many social media platforms continue to operate under a guise of neutrality, refusing to reckon with or even acknowledge their role in actively shaping contemporary culture (Chander & Krishnamurthy, 2018; Gillespie, 2018).

When platform governance fails, social media users are exposed to a variety of sociotechnical harms (Schoenebeck & Blackwell, 2021; Domínguez Hernández et al., 2023). Online harassment, hate speech, and other forms of interpersonal abuse inflict emotional, psychological, and physical distress (Blackwell et al., 2017; Vitak et al., 2017; Im et al., 2022). Rumors, conspiracy theories, and state propaganda distort public understanding and promote radicalization, sometimes culminating in physical violence (Marwick & Lewis, 2017; Starbird et al., 2019; Marwick et al., 2021). Suicide and other forms of self-harm are normalized or even glorified (Chancellor et al., 2016; Pater & Mynatt, 2017).

These harms are compounded by platform designs that perpetuate existing structural inequities, resulting in disproportionate impacts to vulnerable social media users, including women, people of color, queer people, transgender people, refugees, dissidents, and many others (Blackwell et al., 2017; Caplan, 2018; Noble, 2018; Pearce et al., 2018; DeVito et al., 2021).

Platform governance practices further exacerbate these harms, with marginalized users facing disproportionate content moderation experiences (Haimson et al., 2021; Lyu et al., 2024; Mayworm et al., 2024; Thach et al., 2024), due in part to imprecise automated enforcement mechanisms that collapse social and political complexity (West et al., 2019; Gorwa et al., 2020). Despite these disparities, platforms offer users limited opportunities for recourse (West, 2018; Vaccaro et al., 2020).

Technology companies currently wield considerable power over public discourse, with limited accountability for, or transparency into, their specific governance practices (Suzor et al., 2019; Keller & Leerssen, 2020). Though social media platforms have failed to adequately regulate themselves, global regulators have also struggled to effectively respond, in part because platforms, who operate across borders, must navigate a patchwork of local and national laws (Klonick, 2017; Suzor, 2019; Keller & Leerssen, 2020). Suzor (2019) argues that platform governance amounts to private regulation by proxy, often substituting for formal legal mechanisms. As platforms adopt increasingly sophisticated governance regimes, such as Meta's Oversight Board (Klonick, 2019), distinctions between private and public authority blur, further disempowering individual users.

In response to these concerns, novel regulatory frameworks are emerging, though unevenly across jurisdictions. The European Union's Digital Services Act (DSA) and General Data Protection Regulation (GDPR) represent efforts to impose transparency and accountability on social media platforms, particularly regarding content moderation and data usage (Suzor, 2019; Keller, 2022). In contrast, regulatory responses in the United States remain fragmented and largely deferential to corporate self-governance, constrained by First Amendment protections and historical reluctance toward state regulation of speech (Balkin, 2021; Keller, 2023). As social media platforms become increasingly central to contemporary public life, how they are governed—by whom, for whom, and through what mechanisms—remains a critical challenge.

### 1.3.3 Trust & Safety and the "Techlash"

Social media governance has become increasingly professionalized in recent years, largely under the umbrella of "Trust & Safety" (T&S), a common industry term used to describe various people, policies, and products that broadly support user safety. Though the specific origins of the term are unknown, "trust and safety" (and similar variants) have been used throughout the technology industry since at least 1999, with eBay referenced as one early adopter (Boyd, 2002; Cryst et al., 2021).

Contemporary technology companies have largely adopted what Zuckerman and Rajendra-Nicolucci (2023) describe as a "customer service" model of governance, adapting familiar bureaucratic structures for efficient, centralized management of customer concerns to the emergent challenges associated with unanticipated community growth. The increasing professionalization of online governance also emerged in response to growing legal liability, as modern laws—such as the Digital Millennium Copyright Act (DMCA)[3] of 1998—required technology companies operating in the United States to establish formal procedures for responding to potentially infringing content.

Trust & Safety teams, "most often born in a crisis" (Maxim et al., 2022), are formalized over time as companies mature, alongside the evolution from ad hoc governance decisions to explicit policies and processes (Zuckerman & Rajendra-Nicolucci, 2023). While smaller companies may still operate more "artisanal" approaches to platform governance (Caplan, 2018), content moderation practices have rapidly industrialized, with Meta reporting a combined "safety and security" team of 40,000 people (Meta, 2025b). Globally, the Trust & Safety Professional Association (2025) estimates over 100,000 T&S professionals employed in a variety of functions and roles, including policy, operations, and compliance teams, as well as safety-focused roles in product, research, and engineering departments.

The Trust & Safety Professional Association (TSPA)[4], established in 2020 alongside the Trust & Safety Foundation[5] (Goldman, 2020), is one of several recent institutions designed to support, structure, and standardize the T&S profession, including TrustCon[6] (a professional conference first hosted by TSPA in 2022) and the Journal of Online Trust & Safety[7], an academic journal of peer-reviewed research first introduced in 2021 (Cryst et al., 2021). Still, the structure of Trust & Safety teams varies substantially across companies, in part because the governance of a specific platform is informed by its unique characteristics and challenges—but also because T&S functions are often formed extemporaneously, with even founding members of fledgling Trust & Safety teams sometimes lacking relevant prior experience (Maxim et al., 2022; Tyler et al., 2025). This helter-skelter assemblage of professional practice has profound consequences for platform governance, as the present research will demonstrate.

---

[3] Pub. L. No. 105-304, 112 Stat. 2860 (Oct. 28, 1998).

[4] http://www.tspa.org

[5] http://www.trustandsafetyfoundation.org

[6] http://www.trustcon.net

[7] http://tsjournal.org

Recent years have brought significant changes to the burgeoning Trust & Safety industry, most notably following the October 2022 acquisition of Twitter (since rebranded as "X") by Elon Musk, best known for his temporary involvement in the second Trump administration's Department of Government Efficiency (DOGE). During his brief but tumultuous tenure with DOGE, Musk implemented aggressive "efficiency" reforms—including dramatic reductions in staffing and several agency closures—that closely resembled his early management of Twitter, purchased in a $44 billion deal he famously attempted to terminate (Schiffer, 2024). In what some have described as "the most controversial corporate takeover in history" (Mezrich, 2022), Musk immediately cut Twitter's workforce in half—including nearly a third of Twitter's former Trust & Safety team—and reinstated the accounts of more than 6,000 previously banned users, including conspiracy theorist Alex Jones (Basic Online Safety Expectations, 2024; Tyler et al., 2025). In the weeks following Musk's takeover, the average proportion of hate speech on the platform quadrupled (Hickey et al., 2023; Schiffer, 2024).

Many other major technology companies have pursued similar reductions to safety programming in the intervening years, including large-scale workforce reductions—reigniting growing public criticism widely described as the "Techlash" (Su et al., 2021; Helles & Lomborg, 2024). Nearly 760,000 people have been laid off by global technology companies since March 11, 2020—when COVID-19 was first declared a pandemic by the World Health Organization (WHO)—with more than 660,000 of those layoffs occurring since April 14, 2022, the date Musk publicly announced his unsolicited offer to purchase Twitter (Lee, 2020). In January 2025, ahead of the second presidential inauguration of Donald Trump, Meta announced substantial changes to its enforcement policies and products—including ending third-party fact-checking, relocating Trust & Safety teams from California to Texas, and significantly reducing automated content detection and demotion—under the guise of "free expression" (Meta, 2025a). Though company profits continue to soar, technology workers are left demoralized by systemic issues they lack the power to sufficiently address (Su et al., 2021).

*1.3.3.1 Situating platform governance in worker experiences*

While scholars across disciplines have devoted significant attention toward understanding both informal and formal mechanisms for regulating online behavior (e.g., Lampe & Resnick, 2004; Lessig, 2006; Hardaker, 2010; Diakopoulos & Naaman, 2011; Sternberg, 2012; Kiesler et al., 2012; Cho & Acquisti, 2013; Marwick & Miller, 2014; Guberman, Schmitz, & Hemphill, 2016; Massanari, 2017; Blackwell et al., 2018; Im et al., 2022; Han et al., 2023), fewer studies directly examine the experiences and perspectives of the network of professionals—business leaders, government employees, technology workers, and so on—responsible for enacting platform governance at scale.

Scholars who have directly examined platform governance through a worker-centric perspective (Roberts, 2016; Gray & Suri, 2019; Roberts, 2019; Ruckenstein & Turunen, 2020) primarily investigate the experiences of commercial content moderators, or the shadow workforce of outsourced workers responsible for evaluating large volumes of user-generated content in accordance with platform policies and global laws (Roberts, 2019). While understanding the experiences of platform governance's most disenfranchised workers is certainly critical to identifying and advancing opportunities for meaningful change, situating platform governance systems within a broader range of professional practices may produce a more comprehensive understanding of their ideological and operational underpinnings, a necessary prerequisite for systemic transformation.

Direct inquiry of the workers responsible for enacting these technologies is rare, in part, by design. Companies impose stringent (but questionably enforceable) non-disclosure agreements, while individual employees navigate increasing job security concerns and broader threats to their psychological safety amidst frequent, often sudden layoffs (Lee, 2020). As such, most scholarly investigations of platform governance rely primarily on assumptions about professional practices. In the present research, I seek to structure a shared understanding of platform governance across multiple competing perspectives, through direct inquiry of the experiences, perspectives, and interpretations of several categories of differently positioned workers (Suchman, 1995) generally described as *content moderation professionals*. By making this work more visible, I aim to produce a more intimate understanding of the complex landscape of people, practices, and politics that ultimately determines how platforms are governed—which may conflict with companies' platform governance intentions, regulators' platform governance expectations, and the broader public's platform governance assumptions.

Influenced by a growing body of human-computer interaction (HCI) and computer-supported cooperative work (CSCW) literature promoting worker-centric inquiries of technological progress (Irani & Silberman, 2013; Fox et al., 2020; Su et al., 2021; Wolf et al., 2022), this research aims to broaden current understandings of platform governance by examining the lived experiences of practitioners who enact, examine, and engage with these same sociotechnical systems. The results—which both corroborate and complicate existing assumptions about workers' underlying practices—reflect insights from workers with a broad range of both theoretical and practical platform governance expertise, from part-time content moderators and professional researchers to corporate vice presidents.

### *1.3.3.2 Situating platform governance in broader social worlds*

Still, the experiences of individuals—the traditional focus of qualitative inquiry—are only one component of the complex ecosystem of actors that inform and influence contemporary platform governance. Despite technosolutionist promises of orderly, frictionless futures, our social realities are never simple or straightforward, and investigating the "mess" of human complexity (Dourish & Bell, 2011) is necessary to construct realistic paths toward progress. I leverage situational analysis (Clarke, 2005) to ground my interpretation in the broader "situation" or social ecology of social media governance, which necessarily implicates collective actors from multiple social worlds, often with competing interests. By examining the sociopolitical experiences of content moderation professionals in juxtaposition with the institutional processes and power imbalances that construct their discursive realities (Foucault & Nazzaro, 1972), this study seeks to examine the underlying logics that construct and sustain contemporary platform governance and articulate aspirational, worker-centered visions for more equitable technology futures.

Contemporary social media governance is primarily constructed by four social worlds (see Figure 1.1), representing the interests of businesses, states, the public, and the working class. A social world represents a group with broadly shared commitments and practices, each comprising various organizations of actors who participate in platform governance to varying degrees, and with differential access to social and political power. Business interests are the dominant bloc of platform governance, with technology companies themselves holding the majority of power. This is particularly (but not only) true of technology companies with a public-facing social platform product, such as Meta, TikTok, and X, who maintain control over platform design, policy implementation, user data, and technical infrastructure.

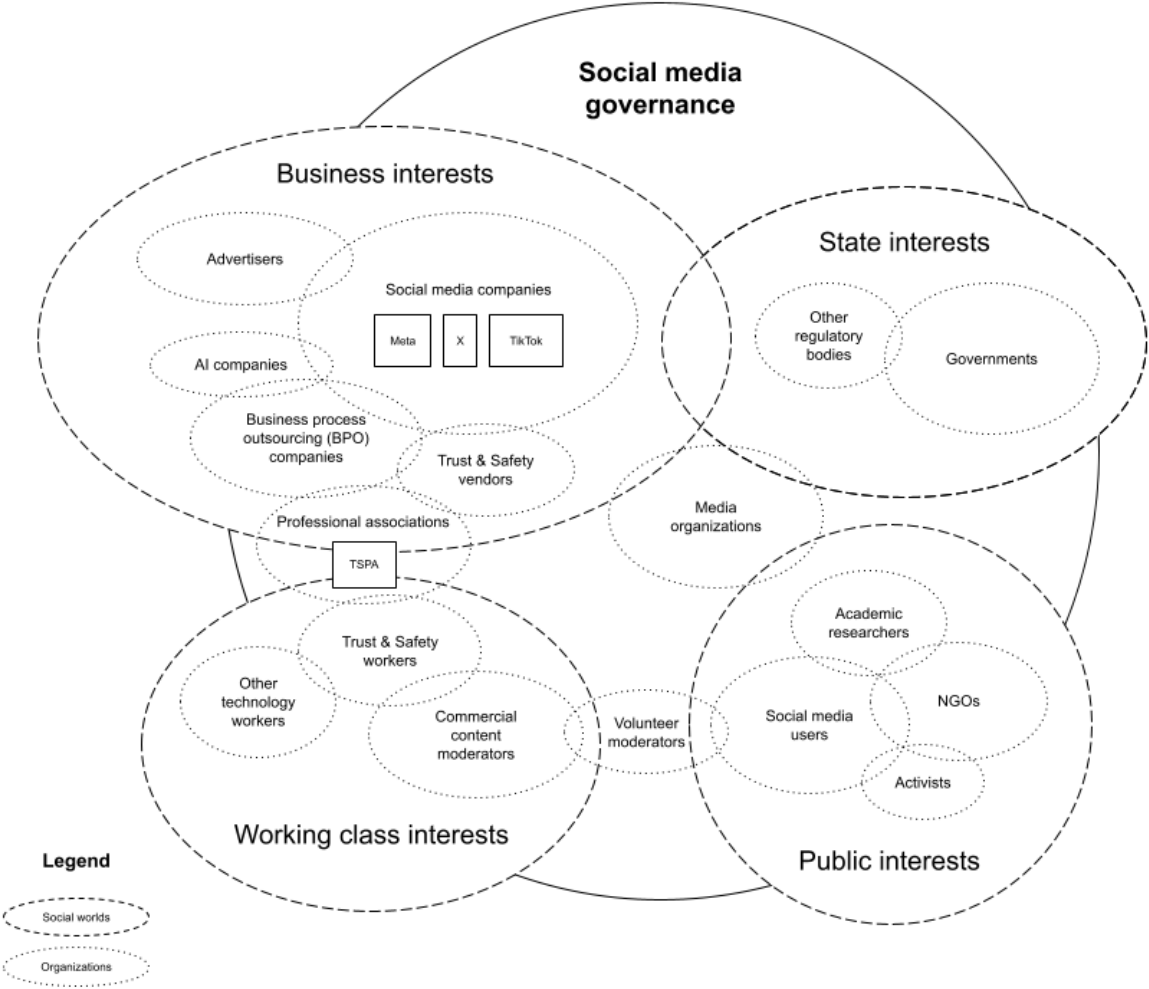

Figure 1.1: A social worlds map illustrating the primary collective actors who construct and maintain contemporary platform governance, including both formal and informal organizations.

Other technology companies also participate in enacting platform governance: business process outsourcing (BPO) companies (such as Accenture, Cognizant, and TaskUs) manage critical content moderation, data labeling, and customer support tasks; artificial intelligence (AI) companies (such as OpenAI, Hive, and Spectrum Labs) develop technologies used to augment or automate content review processes, including machine learning models designed to detect potentially violative or otherwise problematic content; other vendors provide various Trust & Safety services, including content moderation platforms (such as Cinder and Checkstep), identity verification services (such as Persona and Prove), and risk intelligence tools (such as Crisp and CrowdStrike). Also representing business interests are advertisers, who participate in platform governance primarily through their relationships with social media companies, which generate the vast majority of their revenue through targeted advertising.

Businesses are predominantly interested in profitability, in direct conflict with the interests of the working class, whose devalued labor produces profit (Marx, 1844/1959). Working class interests broadly include stable employment, safe working conditions, and fair pay, though class stratification results in privilege and power disparities between workers (Ehrenreich & Ehrenreich, 1977). In the context of social media governance, commercial content moderators—who are typically employed by BPOs in the Majority World, often on a contract basis—are the most structurally disempowered, despite directly managing the daily labor of platform governance. Other technology workers, including those in Trust & Safety and related roles, generally enjoy significantly higher wages and more direct proximity to corporate decision-making. However, as the present research will demonstrate, most technology workers lack the necessary structural power to effectively influence platform governance, confirming the utility of positioning salaried technology employees in broad coalition with their more severely disenfranchised peers. All workers are subject to exploitation and face increasing labor precarity as pervasive automation increases (Riek & Irani, 2025).

State interests, represented by governments and other regulatory bodies, largely influence social media governance through the direct application of formal laws. Though legal requirements vary across jurisdictions, recent regulations include laws governing the visibility of certain content (such as Germany's Network Enforcement Act, 2017); laws protecting user privacy (such as California's Consumer Privacy Act, 2018); and laws mandating corporate transparency (such as the European Union's Digital Services Act, 2022). Certain regulations have outsized influence on contemporary platform governance, such as 47 U.S.C. § 230, a provision of the United States' Communications Decency Act (1996) under which online service providers enjoy limited liability for user-generated content (Gillespie, 2018; Citron & Franks, 2020). State actors may also exert informal or covert influence over platform governance—for example, by pressuring platforms to remove politically sensitive content (Park & Sang, 2023) or sponsoring influence operations designed to manipulate public opinion (Starbird et al., 2019). As in other social worlds, differences in power and proximity to capital determine overall influence, and wealthier or more geopolitically powerful nations may shape platform governance in ways that further marginalize the interests of the Majority World.

Public organizations also inform social media governance, though public interests are generally more dispersed across loosely organized groups of social actors, including activists, academics, and social media users themselves. No organization is a monolith, and the interests of individual social media users clearly vary; however, public interests can be broadly understood as representing autonomy, privacy, and general welfare. Because they lack the institutional power present in other social worlds, public social actors—including people who use, or are otherwise proximate to, technology products and services—are among the most marginalized by existing platform governance structures. While advocacy by non-governmental organizations (NGOs) and other civil society groups can highlight technology harms and recommend potential solutions (Severance, 2013), public actors typically lack structural mechanisms for motivating or otherwise enforcing social progress. Without thoughtful, transparent, and equitable governance, social media platforms risk capture by the interests of powerful and well-resourced actors, rather than serving the interests of the broader public.

## 1.4 Methods

Data was collected during six participatory design workshops held via Zoom in June and July 2022. Participants were not compensated for their participation in this study. Potential participants were recruited via judgment sampling, based on two factors:

1. *Expertise.* Expertise was the primary factor in participant recruitment, as the study required participation from experts who specialize in professional content moderation and platform governance. I specifically recruited experts with a wide range of experiences, including professionals employed in the technology industry and in broader civil society.

2. *Relational trust.* A second factor, particularly for the recruitment of potential industry participants, was a desire to initially contact only those experts with whom the author had some pre-existing relationship, for the safety and comfort of participants as well as the author. Participating in a research study as a representative of a company presents significant risk, particularly in the current corporate technology climate. Industry media coverage increasingly highlights privileged information from presumably internal sources, resulting in fear, distrust, and concern for potential retaliation or job loss. The author did not wish to risk anyone's career, reputation, or material security should someone view participation in this study as a breach of corporate confidentiality.

Recruitment continued until saturation was reached, resulting in a total of 33 participants with a collective 230 years of content moderation experience. Participants' professional content moderation experiences ranged from those working for technology companies (including Airbnb, Discord, Google, Instagram, Lyft, Meta, Microsoft, and Twitter; represented as "I" in Tables 1.1 and 1.2, indicating industry experience) to those working for universities (including Carnegie Mellon University, Cornell University, Georgia Institute of Technology, Northeastern University, Rutgers University, Stanford University, University of California Irvine, University of California Los Angeles, University of California San Diego, University of Colorado Boulder, University of Illinois Urbana-Champaign, University of Maryland, University of Michigan, and Yale University; represented as "A" in Tables 1.1 and 1.2, indicating academic experience) and non-profit institutions (including AI Now Institute, American Library Association, Dangerous Speech Project, Data & Society, and Meedan; represented as "C" in Tables 1.1 and 1.2, indicating other civil society experience). Nine participants had also been content moderators themselves, either professionally or in a volunteer capacity (represented as "M" in Tables 1.1 and 1.2). One participant served as a government advisor (represented as "G" in Tables 1.1 and 1.2).

At the time of recruitment, participants had between 2 and 20 years of experience in content moderation specifically. While the median number of years' experience across all participants is 6, participants who chose to be identified in the resulting manuscript represent more experience on average (median 8 years) than participants who chose to participate anonymously (median 5 years), which suggests that more tenured professionals may be able to discuss their experiences and perspectives more openly than their junior peers. Similarly, most industry participants chose to participate anonymously: of the 16 identified participants, only 7 report industry-specific experience (median 12 years), including 6 former employees of major technology companies (3 of whom were subsequently employed by academic or other non-corporate institutions; one participant was working as an independent consultant, and the remaining two participants were between jobs). Only one identified participant (P8) was employed by a major technology company (Meta) at the time of data collection. In contrast, 11 of the 17 anonymized participants were current full-time employees ("FTEs") of major technology companies (median 5 years). Restrictive non-disclosure agreements and other systemic barriers prevent many workers in the technology industry from openly discussing their experiences and knowledge, even in a research context.

During the workshops themselves, most participants chose to identify themselves to other workshop participants; some chose to participate under a pseudonym. Participants were also afforded the ability to self-identify in the resulting manuscript, at each individual's desired level of identifiability (e.g., with or without a name or institutional affiliation). While participants were invited to participate in the workshops with or without video, all participants were required to join from a laptop or other device that would allow them to view a shared screen. Participants were also given links to workshop documents (e.g., Google Slides), should they wish to contribute directly to the documents themselves. Relevant documents were secured after each workshop concluded. Workshops were recorded (voice, screen, and chat) and transcribed for research purposes only. Participants were able to change their level of identifiability or withdraw from the study at any time before the paper's publication.

Two workshops had seven total participants; two workshops had four participants; one workshop had five participants; and the final workshop had six participants. In order to more appropriately examine the broader social ecology of platform governance, each workshop included content moderation professionals from various social worlds, to surface the perspectives of diverse individuals while also facilitating interactions between them. Workshops were structured as a *future workshop*, a participatory design method used to collaboratively define an ideal future outcome, without existing resource restraints, technical limitations, and organizational realities (Jungk & Müllert, 1987; Vidal, 2006; Hardy et al., 2022). After orienting around a collective vision for a better future, participants determine potential solutions and procedures to advance the status quo closer to this imagined future. The goal of a future workshop is to bring together a diverse group of people who share interest in a common problem—in this case, scaled content moderation. The structure of a future workshop allows participants to identify both shared and conflicting goals, which can then be used to structure a collective path forward. A future workshop has three stages: 1) *Inspiration* (identify common problems; e.g., reflect on the present-day situation), 2) *Ideation* (generate a shared vision for a better future; e.g., what would the ideal solution be, assuming anything is possible?), and 3) *Implementation* (discuss and prioritize potential ideas and solutions).

Table 1.1: Identified participants (self-described), including academic (A), other civil society (C), government (G), technology industry (I), and moderator (M) experiences at time of data collection (2022).

|     | Name                   | Role(s)                                                                                                                              | Experience          |
| --- | ---------------------- | ------------------------------------------------------------------------------------------------------------------------------------ | ------------------- |
| **P1**  | Susan Benesch          | Executive Director, Dangerous Speech Project                                                                                         | 9 years (C)         |
| **P2**  | Robyn Caplan           | Senior Researcher, Data & Society Research Institute<br>*(formerly)* Researcher, Rutgers University                                  | 6 years (A,C)       |
| **P3**  | Anne Disabato          | *(formerly)* Product Manager, Facebook<br>*(formerly)* Product Manager, Twitter                                                      | 7 years (I)         |
| **P4**  | Casey Fiesler          | Associate Professor, University of Colorado Boulder<br>Moderator                                                                     | 5 years (A,M)       |
| **P5**  | Eric Gilbert           | Associate Professor, University of Michigan<br>*(formerly)* Associate Professor, Georgia Institute of Technology                     | 7 years (A)         |
| **P6**  | Sarah Gilbert          | Research Manager, Citizens and Technology Lab, Cornell University<br>*(formerly)* Researcher, University of Maryland<br>Moderator    | 6 years (A)         |
| **P7**  | Amber Grandprey-Shores | Moderator, Twitch<br>Moderator, Discord                                                                                              | 5 years (M)         |
| **P8**  | Mark Handel            | Researcher, Meta                                                                                                                     | 10 years (I)        |
| **P9**  | Del Harvey             | Independent consultant<br>*(formerly)* Vice President, Trust & Safety, Twitter                                                       | 20 years (I)        |
| **P10** | Matthew Katsaros       | Director, Social Media Governance Initiative, Yale University<br>*(formerly)* Researcher, Twitter<br>*(formerly)* Researcher, Facebook | 11 years (A,I)      |
| **P11** | Kat Lo                 | Content Moderation Lead, Meedan<br>*(formerly)* Researcher, Instagram<br>*(formerly)* Researcher, University of California Irvine<br>Moderator | 14 years (A,C,I,M)  |
| **P12** | Sarah T. Roberts       | Associate Professor, University of California Los Angeles<br>*(formerly)* Staff Researcher, Twitter<br>*(formerly)* Fellow, American Library Association | 12+ years (A,C,I)   |
| **P13** | Joseph Seering         | Researcher, Stanford University<br>*(formerly)* Researcher, Carnegie Mellon University                                               | 7 years (A)         |
| **P14** | Jan Smole              | *(formerly)* Trust & Safety Process Manager, TikTok<br>*(formerly)* Community Operations, Facebook                                   | 13 years (I)        |
| **P15** | Kristen Vaccaro        | Assistant Professor, University of California San Diego<br>*(formerly)* Researcher, University of Illinois Urbana-Champaign          | 8 years (A)         |
| **P16** | Sarah Myers West       | Managing Director, AI Now Institute<br>Senior Advisor, Federal Trade Commission                                                      | 8 years (C,G)       |

Table 1.2: Anonymized participants (self-described), including academic (A), other civil society (C), government (G), technology industry (I), and moderator (M) experiences at time of data collection (2022).

| | Role(s) | Experience |
|---|---|---|
| **P17** | Associate Professor<br>Moderator | 5 years (A,M) |
| **P18** | Lecturer, Northeastern University<br>Researcher, Stanford University<br>Moderator | 2 years (A,M) |
| **P19** | Researcher<br>Moderator | 5 years (A,M) |
| **P20** | Research Scientist | 3 years (A) |
| **P21** | Researcher, Twitter<br>*(formerly)* Researcher, Facebook | 6 years (I) |
| **P22** | Researcher<br>Program Manager | 4 years (A,C) |
| **P23** | Researcher | 5 years (A,I) |
| **P24** | Researcher<br>*(formerly)* Researcher, Meta | 3 years (I) |
| **P25** | Research Manager, Meta<br>Researcher, Meta | 4 years (I) |
| **P26** | Senior Designer, Game Company | 10 years (I) |
| **P27** | Data Scientist, Google<br>*(formerly)* Data Scientist, Facebook | 7 years (I) |
| **P28** | Data Science Manager, Twitter<br>Data Scientist, Twitter | 2 years (I) |
| **P29** | Product Manager, Meta<br>*(formerly)* Product Manager, Microsoft | 4 years (I) |
| **P30** | Policy Manager<br>*(formerly)* Trust & Safety Project Manager, Facebook<br>*(formerly)* Market Specialist, Facebook | 4.5 years (I) |
| **P31** | Program Manager, Community Safety, Discord<br>*(formerly)* Trust & Safety Specialist, Lyft<br>*(formerly)* Trust & Safety Specialist, Discord<br>*(formerly)* Trust & Safety Specialist, 24/7 InTouch at Airbnb | 7.5 years (I,M) |
| **P32** | Journalist, Washington Post | 5 years (C) |
| **P33** | Data Scientist, Twitter<br>*(formerly)* Researcher, Facebook | 5+ years (I) |

To facilitate this collaborative problem-solving, workshops began (following a brief introduction) with a virtual adaptation of the KJ method of brainstorming (Kawakita et al., 1967; Scupin, 1997). Participants were asked to independently identify obstacles to successfully moderating content at scale, using Google Jamboard (a virtual brainstorming tool). After collaboratively grouping similar ideas under larger themes, each participant noted their top three priorities, or the most important obstacles to resolve, resulting in a categorized set of collective priorities. Next, participants engaged in a paired journey mapping exercise to more tangibly illustrate their understanding of current scaled content moderation processes. After each pair shared their journey map with the larger group, participants were asked to reimagine their maps using the magic wand approach, temporarily setting aside any practical obstacles (e.g., resource constraints). Finally, participants collaboratively generated a roadmap to move us closer from the status quo to their collectively imagined future, populating a feasibility grid with key objectives—i.e., what can be accomplished in the next year? 5 years? 10 years?—with a focus on specific tasks and outcomes.

I conducted a thematic analysis (Braun & Clarke, 2006) of the resulting data, which included voice and chat transcripts as well as visual artifacts collaboratively produced by participants during each workshop. I used an inductive approach to develop codes, individually reading workshop transcripts and noting codes by hand. After discussing these initial codes with a research assistant, I created a more comprehensive list of codes (36 codes in total). I manually coded two transcripts in a pilot coding process to test and refine the codebook. Resulting codes were organized around several themes, including but not limited to cultural context; issues of scale; and corporate values and practices. Quotations have been lightly edited for readability.

I also draw from methods of situational analysis (Clarke, 2005), utilizing social world mapping (see Figure 1.1) to visualize the influences of and relationships between the collective actors involved in negotiating contemporary social platform governance. Situational analysis provides a framework for interpreting how different groups engage in complex negotiation, facilitating analysis of competing interests and power differences. My analysis draws on multiple intersecting data sources, including workshop transcripts, visual artifacts, and other discursive materials, as well as my own experiences in the professional content moderation industry.

*1.4.1.1 Position statement*

My research questions, analysis, and contributions cannot be understood independently of my position as the researcher, which includes my experiences as a white, cisgender woman living in the United States (Williams & Irani, 2010; Bardzell & Bardzell, 2011). This particular study and its methodological approach are made possible by my unique position as both an academic researcher and an industry practitioner, which affords me a unique level of familiarity with the inner workings of major technology companies. In addition to being a PhD candidate at the University of Michigan (where I have conducted social media research since 2014), I have been directly employed by several different technology companies, beginning with Meta (then Facebook) in 2017. I have also been employed by Twitter (now X) and two social media startups (Sidechat and YikYak), and I currently lead Trust & Safety operations at Mozilla. While the present research is not an ethnographic inquiry, it is necessarily informed by years of professional research about, and applied practice in, social media governance. My research is also informed by my experiences as a queer, disabled person and technology worker, and by my personal experiences with online harassment and other abusive behaviors technology companies typically wish to prevent on their platforms.

*1.4.1.2 Limitations*

This research represents perspectives from participants who are largely (though not exclusively) Western, white, and highly educated, reflecting broader trends in the technology workforce (U.S. Equal Employment Opportunity Commission, 2024). Experiential understanding from more precarious technology governance workers is also notably absent from the current work, due to the risks associated with study participation as well as the limitations of my own professional network. Future research should investigate these additional perspectives, including those of workers located in the Majority World.

## 1.5 Results

The results are organized into three sections: content moderation foundations (*what are we trying to achieve?*), content moderation realities (*where are we falling short?*), and content moderation futures (*what can be done?*).

***Content moderation foundations.*** What are we trying to achieve? Across social worlds, participants agreed that successful content moderation is *principled, consistent, contextual, proactive, transparent,* and *accountable.* First and most critically, participants emphasized the impossibility of "neutral" governance, urging platforms to commit to specific values instead of striving for the unattainable goal of universal applicability. Efforts to appear impartial yield vague, imprecise policies that cannot be operationalized at scale—but like other forms of governance, successful moderation requires rules to be consistently enforced, in order to establish clear expectations for appropriate behavior. Successful moderation also accounts for relevant context, though this is difficult (if not impossible) to achieve at scale, given both the volume of content users produce and its global context. Given the challenges associated with scaling reactive moderation, participants advocated for more proactive—and preventative—approaches to mitigating harm. Finally, participants stressed that successful moderation is both transparent and accountable: users must understand how and why decisions are made, with meaningful opportunities for recourse when mistakes inevitably occur. Transparency and accountability not only build trust but also confer legitimacy, encouraging compliance even when users disagree with individual policies or outcomes.

*Content moderation realities.* Where are we falling short? Participants described numerous barriers to achieving these goals on contemporary social media platforms, including the impractical pursuit of "one size fits all" solutions to global governance—resulting in vague, Western-centric policies that ignore cultural nuance and further marginalize non-dominant groups. These structural flaws are compounded by the demands of scale, as machine learning models can only reliably detect content with broad agreement. Despite these challenges, many technology companies pursue what participants described as a "growth at all costs" business model, creating strong incentives that frequently and directly conflict with user safety goals. Participants characterized company leaders as reluctant to invest in content moderation and other work related to user safety, often prioritizing new product development over the mitigation of existing product risks. Despite limited resources, workers are expected to support continual platform expansion and scaled growth, resulting in what participants described as an overreliance on third-party labor and other exploitative working arrangements. Content moderation labor and other data labeling tasks are outsourced to vendor workforces in regions with lower labor costs, while civil society organizations engage in unpaid labor they may never see implemented. Individual workers are left feeling unimportant, overburdened, and isolated from their closest peers, resulting in what participants describe as chronic burnout and frequent employee turnover.

*Content moderation futures.* What can be done? After identifying numerous ways in which contemporary technology companies fail to successfully govern their platforms, participants considered potential solutions. First, participants encouraged company leaders to clearly articulate a holistic vision of community safety—one that recognizes content moderation as an integral component of the product experience, rather than a crisis response. Participants highlighted the role of platform design in shaping user behavior, advocating for designs that establish clear expectations for appropriate conduct and promote the rehabilitation of potential offenders. Because existing corporate structures have failed to adequately protect users' safety, participants stressed the need for greater corporate accountability, whether incentivized via regulatory structures or alternative levers. While participants expressed support for the development of new regulations to force more responsible corporate action, they questioned lawmakers' proficiency in a problem space rife with complexity and nuance—and they cautioned that technology companies may be incentivized to identify loopholes to avoid increased scrutiny. Finally, given the challenges associated with uniform governance of general-purpose platforms, participants advocated for more localized and participatory alternatives, empowering users to shape and steward their own community spaces.

Table 1.3: Overview of results.

| **Content moderation foundations:** What are we trying to achieve? | | |
|---|---|---|
| Successful moderation is **principled** | *Platforms must define their values; neutrality is impossible* | p. 21 |
| Successful moderation is **consistent** | *Consistent enforcement creates clear expectations* | p. 22 |
| Successful moderation is **contextual** | *Accurate, equitable outcomes consider relevant context* | p. 22 |
| Successful moderation is **proactive** | *Instead of reacting to harm, prevent harm from occurring* | p. 23 |
| Successful moderation is **transparent** | *Users should understand how and why decisions are made* | p. 24 |
| Successful moderation is **accountable** | *Avenues for recourse cultivate trust, process legitimacy* | p. 25 |
| **Content moderation realities:** Where are we falling short? | | |
| **One size fits none:** The futile pursuit of universal agreement | *Lack of industry consensus restricts cohesive progress* | p. 27 |
| | *When global rules are defined locally, Western perspectives dominate* | p. 28 |
| | *Striving for universal applicability dilutes policy effectiveness* | p. 29 |
| **Problems of scale:** "Lowest common denominator" moderation | *Crude solutions can't be scaled* | p. 30 |
| | *Scale flattens nuance* | p. 31 |
| **Growth at all costs:** Misalignment with business incentives | *Business incentives undermine safety goals* | p. 32 |
| | *Metrics obsession drives optimization of numbers, not outcomes* | p. 33 |
| | *Flawed "innovations" prioritized over fundamental safety* | p. 34 |
| | *Preoccupation with short-term costs prevents long-term gains* | p. 35 |
| | *Constant deprioritization results in chronic underinvestment* | p. 36 |
| **Labor pains:** Exploitation, isolation, and burnout | *Overreliance on third parties and exploitative working arrangements* | p. 38 |
| | *Fractured understanding: Navigating organizational silos* | p. 39 |
| | *Burnout, turnover, and institutional knowledge loss* | p. 40 |
| | *Executives enjoy disproportionate influence* | p. 42 |
| **Content moderation futures:** What can be done? | | |
| Embrace **holistic visions** | *Moderation is essential, forever, and communal* | p. 44 |
| Design platforms that **encourage prosocial behavior** | *Establish clear expectations for appropriate behavior* | p. 47 |
| | *Promote rehabilitation, not retribution* | p. 49 |
| | *Favor flexible interventions* | p. 50 |
| | *Implement responsive regulation* | p. 51 |
| Incentivize **corporate accountability** | *Regulation: Promising, but not a panacea* | p. 53 |
| | *Leverage alternative accountability mechanisms* | p. 55 |
| Explore **alternative models** | *Forsake the 'killer app'* | p. 57 |
| | *Empower community governance* | p. 58 |

**1.5.1 Content moderation foundations: What are we trying to achieve?**

A foundational challenge identified by many participants is the lack of a unified, strategic vision for what content moderation should ideally achieve, particularly at the scale afforded by modern social media platforms. P9 said:

> "Obviously, one of the biggest challenges is that there actually isn't a real definition of what successfully moderating social media at scale looks like in the first place. What is 'success' in that context? What does it mean for it to be successful?"

*1.5.1.1 Successful moderation is principled*

First and most critically, participants noted the importance of establishing foundational principles to guide moderation practices. "User perceptions of harm can be complex," said P23. "But in a content moderation context, I think platforms have to take a stand. Basically, it becomes a question of: are you going to take the user-centered approach? What's the cost that comes with that? If there's one group of users who think child sexual abuse material is completely okay, are you going to listen to them?"

A failure to establish consistent values—often in favor of something resembling "neutrality"—can instead reinforce existing systems of social oppression, such as racism and transphobia. P24 said: "Some criticism I've heard, from within and outside of the company, is that a lot of the policies are written in a perspective that favors certain groups. There's definitely been a lot of criticism of how people of color are treated, how LGBTQ+ people are treated, how overweight or plus-size people are treated, and how the policy tends to favor cis white folks." P7 agreed:

> "You see that a lot with content creators. I follow streamers who are activists as well, and some of them have actually had conversations with companies like Twitter and been like 'Yes, I was in fact shadow-banned, because I was being loud on main.' And then you see folks talking about 'I can't swear, I can't talk about being black, I can't talk about being gay, I can't talk about this, that, or the other thing,' which is a problem."

Instead, many participants emphasized that neutrality is not an achievable goal. P32 said: "There is no global set of values. At some point, you just stop trying to please everybody and make a choice. What are *your* values?"

### 1.5.1.2 Successful moderation is consistent

Attempts to remain impartial can produce ambiguous policies that are difficult or even impossible to successfully operationalize at scale, resulting in inconsistent and imprecise moderation. Successful moderation, like other forms of governance, requires rules to be consistently applied. "Moderation needs to be active and continuous in order for internal norms to be maintained," said P27. Inconsistent moderation—whether due to middling principles or unreliable enforcement—results in unclear expectations for user behavior. P27 continued:

> "If we don't have clarity around what the rules are within a particular community, people will be more inclined to break those rules due to lack of knowledge or lack of clarity as to what is considered acceptable."

Inconsistent moderation also creates additional opportunities for malicious actors, who may be willing to risk potential enforcement when consequences are not guaranteed. "Consistent response is important," said P33. "Inconsistent, slow enforcement creates space for adversarial users to think, 'It's worth it.' A scammer with three days to scam someone is likely to think it's worth the effort."

### 1.5.1.3 Successful moderation is contextual

Effective moderation also considers context, which becomes increasingly challenging at scale, given both the volume of content users produce and its global context. For individual moderators, additional time spent reviewing particularly challenging or otherwise ambiguous content may jeopardize the metrics by which their performance is strictly evaluated—further complicated by standardized review tools that may not display all potentially relevant information. "When you are on the ground doing this kind of evaluation, it can be really challenging," said P6. "Even if you have a really solid policy, it can be challenging to make correct determinations based on that, because there's so much missing context—context that you might not actually have access to. It also takes time, and you have to respond to some of these issues really, really quickly."

Even without the constraints of time and tooling, context can be difficult to glean; language evolves quickly, and online interactions occur in diverse social contexts. "A lot of harms are referenced by coded language—and the coded language can evolve faster than people can detect it," said P8. "Also, so much of this is situated within the context of friendship. It's sometimes hard to tell the difference between friendly banter and true harm." While it isn't realistic to expect all relevant information to always be available (or accessible), providing moderators with as much context as possible is critical to producing accurate and equitable governance outcomes.

*1.5.1.4 Successful moderation is proactive*

Many participants highlighted the industry's focus on reactive moderation practices, which P9 described as "an over-focus on interventions being at the point where harm has already been done in some way—versus attempting to intervene potentially before that point." P10 agreed:

> "Most moderation is happening far too late. More proactive and upstream work is rarely prioritized or thought about. Often we're talking about the decision that happened, but not the days, months, weeks, years that led up to that point in time."

Participants strongly favored more proactive strategies, such as improved user education and other risk mitigation efforts designed to prevent at least some policy-violating content from ever appearing on the platform. P7 said:

> "I've worked in operations for a very long time. We want to react; we don't proactively create systems that can address some of the problems. You can't catch everything ahead of time—there is no way to predict a violation before it happens—but you can have systems in place. Twitter has the thing where it's like, 'Are you sure you want to send that?' Interference and friction is a good way to stop things from happening."

Part of this over-reliance on reactive moderation is a function of underinvestment, with many companies only turning their attention and resources toward platform governance once some issue has already arisen. P5 said:

"I have seen with a lot of start-ups something you might call the 'Oh Shit' model, where very little if anything is ever getting moderated. The first real moderation actions come in reaction to some personal referral. Somebody knows somebody at the company, an engineer or PM or executive... I've seen this happen in at least two start-ups, and I wouldn't be surprised if it's most start-ups."

To the extent that platforms do invest in proactive mitigations, it is typically natural language processing algorithms designed to detect potentially violative content before a user reports it. "'Proactive' in industry tends to mean getting rid of something before it's reported," said P13. "To me, proactive means something much more like preventative care in healthcare—the issue was prevented."

### 1.5.1.5 Successful moderation is transparent

Participants widely agreed that successful moderation is transparent, and that a lack of transparency erodes user trust. P27 said:

"When active moderation takes place, if there is a lack of transparency in the decision-making process, that can lead to frustration and ferment dissatisfaction and fracturing in the community. A successful community is one where when hard decisions are made, there is clarity and transparency around the final outcome. Any successful community needs to have a transparent moderation process in place."

Several participants referenced platforms' belief that increased transparency will lead to undesirable outcomes, as "bad actors" may be better positioned to manipulate their systems. "Companies are reluctant to be transparent about their moderation processes because they are afraid of the bad actors who game their systems," said P20. Instead, many platforms restrict what they share with the public, often obscuring important details and fracturing users' understanding of what is or is not allowed. P20 continued: "The community rules are usually vague, so different users have different interpretations of the rules."

In the absence of meaningful transparency, users are left confused by moderation decisions they do encounter on the platform; they may not even recognize that moderation has occurred. "I don't think content moderation can ever be called successful when most users have no knowledge of it," said P1. "And no agency over it. At all." P4 agreed, suggesting that users who do engage in policy-violating behavior are "not being told what the problem is":

> "TikTok's moderation system has some problems. One of them is a severe lack of transparency when it comes to what rule has been violated—why someone's content is being removed, or why someone has been banned. I've been told by other creators that when you hit the appeal button, there used to be a 'Tell us why you're appealing.' Now, there's not. It's just, 'This content was removed because of a Community Guidelines violation.' You click a button to appeal, and what happens after that, you don't know. At some point in the future, you either get something back that says 'We reviewed your content, and it was restored'—and you still don't know what they think you did wrong— or 'We reviewed your content and decided it was a violation,' and you don't know what you did wrong and whatever rule you broke. You don't know how to not break it again."

Reduced transparency also limits the extent to which content moderation experts and the general public can provide feedback to platforms, hindering potential improvements to their moderation systems and practices. "This lack of defensible and extremely transparent governance practices restricts the ability for companies to receive feedback," P25 said, noting that platforms should publicly outline "how our systems view and handle each type of problematic content and solicit feedback from experts on improving the status quo."

### *1.5.1.6 Successful moderation is accountable*

Finally, participants agree that successful moderation practices must have mechanisms for ensuring accountability, both to users and the broader public. Accountability is critical not only for external purposes (such as meeting evolving regulatory requirements), but because fully accurate content moderation is simply not possible, particularly at scale. When mistakes inevitably occur, accountability mechanisms such as appeals processes afford users some amount of recourse. P4 said:

> "Errors and automated moderation… the combination of these two things is really bad. Like, if you're going to have errors—which, I mean, you're going to—especially if they might be systemically biased, which we also see some evidence of on TikTok, then you have to have a good appeals process."

As with offline systems of governance, accountability is paramount to ensuring user trust. "A big obstacle to successfully moderating at scale? Process legitimacy," said P17. "You could have the most effective triage process, but if people don't believe in it producing fair outcomes, then they're just going to undermine it. Looking at recent Supreme Court decisions, we see that playing out in different ways." P6 agreed that accountability is critical in building trust between a platform and its users—or, conversely, that a lack of accountability will have the opposite effect:

> "Building trust in the system, between users, and the policies, and how they're implemented… that is especially important now, where we've had a number of scandals—across a number of different platforms—that have all resulted in eroding trust between users and platforms."

Participants also emphasized the distinction between governance legitimacy and endorsement, noting that even users who disagree with individual policies may still perceive them as legitimate if the process of creating and enforcing those policies is believed to be fair. "People don't have to all agree with all the decisions," said P29. "But they have to respect the process by which we made them, and they have to respect where we have chosen to draw the line." P26 agreed: "I think process legitimacy, depending on the circumstance, can be really important—but as long as you're in that window of legitimacy, it doesn't have to be super well-received. It just has to be something that people can't really push back on." This perspective aligns well with *procedural justice* (Tyler, 2006), which suggests that people are more willing to comply with rules when they perceive the organizations responsible for their enforcement as legitimate authorities (Tyler, 2007; Katsaros et al., 2022). P29 continued:

"If you can sort of get to the point where, through the combination of process and outcome, you have something that is sufficiently defensible, then hopefully, you can get at least the good faith actors in this ecosystem—which, granted, a lot of the people who yell very loudly are not—to the point where they understand the decisions, respect the decisions, are willing to follow the decisions… even if they're not necessarily willing to say, 'Yes, this is something I would stand behind or agree with.'"

In the absence of global consensus, process legitimacy functions as a stabilizing force—but as participants emphasized, platforms must deliberately cultivate institutional trust to maintain governance authority. "Efforts to affect the *efficacy* of moderation are often orthogonal to efforts to affect *perception* of moderation," P26 continued. "If you make your detection better, oftentimes that doesn't pay it forward in terms of people thinking that you're making better decisions. You have to work on both."

**1.5.2 Content moderation realities: Where are we falling short?**

While participants agreed on what characterizes successful moderation at scale, they also described numerous ways in which contemporary platforms are falling short of stated goals.

*1.5.2.1 One size fits none: The futile pursuit of universal agreement*

While participants overwhelmingly agree that successful content moderation is both principled and consistent, most participants characterized current moderation practices as inconsistent at best—in large part due to the difficulty (if not impossibility) of creating policies, protocols, and other governance mechanisms designed for global application.

***Lack of industry consensus restricts cohesive progress.*** Participants expressed frustration with what they perceived as a lack of industry standards, preventing consistent progress across company lines. "There's no general consensus among experts or users on what is an adequate system," said P25. "The conversation is always evolving. What is overreaching? What is under-supporting users?" Even key terms used across the industry, such as *online harassment*, are not consistently defined. "How feminists define harassment can be very different than the people using #Gamergate," said P4. "I always remember a quote from a paper where they interviewed people who identified as part of Gamergate. They were like, 'Oh, that's not harassment. That's just words.'"

For many participants, this lack of standardization across the content moderation industry—or even within a single company—results from the inappropriately monolithic treatment of large, global userbases. "There's this vision of a 'free speech town square' for big social media platforms, versus a community with agreed norms," said P14. "I think it's a core issue. As much as these companies stress 'community' in public statements, there isn't really a community, or any agreed norms for that community." P9 agreed: "Millions of people are a bit too big to just be one community, with shared values and beliefs and principles and the like. Treating all the different communities that exist and interact with each other as somehow one giant community almost inevitably leads to pain."

***When global rules are defined locally, Western perspectives dominate.*** In the absence of global standards, companies must define their own—resulting in governance policies and processes dominated by largely Western perspectives. "Part of what makes a lot of the cases we see in this space much worse is when our systems are imagined by and encode a very Western, cisgender, white, straight perspective into the decisions that are being made," said P26. P7 agreed: "If you have a room full of cishet white men who are all 50 and up, you have a problem. Having centralized policies and guidelines for moderation is very important—but they're only very good and very helpful if you have a lot of voices in the room, all being heard."

P32 referenced this "lack of cultural awareness" as a major obstacle to successful scaled moderation. "The majority of policies are written by people in the United States—and they apply to people all over the world," P32 said. "They're not even in their languages. Considering that most users are actually outside the United States, it's just a fundamentally broken piece of the system." P21 agreed:

> "There's a lack of cultural context regarding harassment, abuse, intimate imagery, toxicity, privacy… what is shame? What is embarrassment? I've done a lot of work in non-Western countries, and what privacy or intimate imagery is in Saudi Arabia is not what it is in Finland. When it comes to content moderation, we really run into a lot of problems because we can't have country- or culture-specific policies."

Participants emphasized that content policies, moderation processes, and other product experiences do not adequately account for the experiences of marginalized people. "I come from the activist side," said P22. "What we're seeing is an interest in viewing the experiences of marginalized groups—people who are not necessarily, you know, privileged users of whatever tech platform—as edge cases, rather than the baseline. That's a huge issue, especially when we think about refugees, trans women, LGBTQ individuals in general, women's rights… all of these things. We are always kept on the margin, and dealt with—for policy application as well as remediation—as edge cases, which creates a lot of issues. At the end of the day, there is no context applied, even though a lot of tech companies say that they engage with stakeholders from different regions."

***Striving for universal applicability dilutes policy effectiveness.*** Though some amount of standardization is required to progress scaled solutions, participants emphasized that universal agreement simply is not possible. "You are looking for agreement on a taxonomy," said P15. "What is toxic? What should be removed? What you're asking for is a world without humans. No matter how toxic something is, there will still be some disagreement. And as soon as you get something that's more 'borderline,' there's going to be a whole lot of disagreement."

While companies could choose to adopt a more principled stance, they hope to attract the broadest possible customer base—resulting in ineffective, overly generalized policies that lack the specificity and coherence to be effectively implemented at scale. "Policies are unclear, ever-changing, reactive, and overwhelming in amount and detail," said P14. "This super granular policy that dances around the issues, rather than a proper set of norms that the community agreed to by joining, is a massive problem—with a lot of problems coming downstream from that." P17 agreed: "Different identity groups will have very different points of view on what toes versus crosses the line. Platforms end up making these very decontextualized policies that apply across many different application areas, and thus really piss a lot of people off."

Ultimately, participants recognized necessary trade-offs between content moderation practices that more appropriately account for individual differences and those that efficiently scale. P1 said:

> "The dilemma at the heart of this is that modern moderation at scale requires some universal agreement—but we also want to honor all of the cultural differences and other differences among people, and that seems to make doing anything at scale especially impossible. You could establish universal definitions—as we have for lots of things, especially in international law—and then do moderation differently in different cultures and different contexts, in order to respect those different cultures and contexts and the values of those users. But, as somebody pointed out earlier, sometimes users have what we consider terrible ideas. So it's extremely tough, you know, balancing opposing input and essentially imposing the ideas of, let's face it, over-educated cultural elites from a few countries on lots of other people—the global majority."

*1.5.2.2 Problems of scale: "Lowest common denominator" moderation*

Despite this lack of standardization, content moderation processes must still be scaled—resulting in crude solutions that predominantly address issues with broad consensus, at the expense of more nuanced concerns.

***Crude solutions can't be scaled.*** While many companies rely on machine learning models to automatically detect potentially violative content, the performance of these models is dependent on the quality of the data used to train them. Accurate detection, for example, relies on clear, consistent definitions—the very consensus still lacking in many safety-adjacent problem areas. "From some of our own work, we know that in one of the biggest datasets, even the stuff that's rated as 'toxic' has like a third of annotators disagreeing," said P17. "It's just a very highly disagreed upon area." P26 agreed: "How do we label something as 'toxic?' That can be super subjective."

Participants reflected on the inherent limitations of attempting to automate detection of content that even humans cannot consistently identify. "There's a really interesting paper by Mitchell Gordon called 'The Disagreement Deconvolution,'" said P13. "He wrote it for a machine learning audience, but he's basically quantifying the argument that reasonable people disagree. There's a kind of peak to model performance where we can't get better—because there isn't objective truth above a certain level." Even for more straightforward types of content, any classification task is subject to certain limitations, such as the inverse relationship between *precision* (e.g., of all emails marked as spam, how many actually contained spam?) and *recall* (e.g., of all actual spam emails received, how many were detected?). P28 noted that optimizing precision and recall is only more challenging when attempting to automate content detection in complex problem areas:

"When we speak about moderation at scale, the easiest way to scale—and the cheapest way—is through algorithms and AI. But all algorithms, all machines, have precision-recall tradeoffs: they all have some false positives and false negatives. And the more challenging a problem space is, the higher chance that precision-recall tradeoff will be suboptimal—it's really hard to achieve good precision-recall in a difficult space. So those most difficult cases end up manually moderated, because algorithms are not detecting them properly. And now a human needs to make a judgment, whether the content is in violation of policies—but again, this is the most difficult type of content, so there is a lot of subjectivity and interpretation of guidelines. AI cannot really help with this. This is where it reaches its prediction power plateau."

***Scale flattens nuance.*** Several participants noted the ways in which scale inherently minimizes difference, resulting in moderation practices that cannot appropriately consider context. "Automated tools flatten behaviors," said P19. "They get more abstract, and then you start to miss local context."

Participants emphasized the difficulty of consistently enforcing policies across communities with divergent norms. "It's very hard for decisions to strike the right balance between nuance and consistency," said P29. "The inherent tension is that, for a really clear process, you want something that's relatively clear-cut—but then it runs up against, like, real life, and the fact that communities do have different norms." P26 agreed:

> "One of the challenges I see is folks sort of colliding in spaces, with very low context for each other and coming from different groups, or even just different circumstances—like 'Hey, I'm having a bad day; I'm not resilient to somebody being mean to me.' Simplifying that basically means that the average, the default, the lowest common denominator moderation does not set anybody up for success in those interactions, however you define success. And as a result, a lot of our systems either flatten or discourage expression of different identities—or make it so that spaces are differently friendly to particular cultures or groups. Sometimes that's desirable; sometimes it's not."

Instead of reckoning with this complexity, many platforms focus their attention on moderating less nuanced violations. "Most platforms wind up with very, very weak table stakes because they don't like the fact that there's all of these diverse perspectives and disagreements," said P17. "We end up only moderating the stuff that is sort of universally agreed upon—like, a racial epithet, that should definitely go. The stuff that is more in the gray area tends to not be moderated, or left up to the communities themselves."

When complex human interactions are evaluated in accordance with decontextualized rules designed for scaled enforcement, predominant norms prevail—often reproducing global power asymmetries. P6 reflected on the reinforcement of dominant experiences at scale:

> "In the community that I moderate, sometimes we don't know. Like, the use of the C-word—totally unacceptable in the U.S., totally normal in Australia. So what do you do when somebody is using that? Do you just reinforce an already American hegemony of platforms and technological spaces by getting rid of that? Or do you potentially have this word that is interpreted by a lot of people as misogynistic? There's all these trade-offs that have to do with power that are really challenging to scale."

Participants noted the ways in which this 'flattening' of individual expression risks further marginalizing vulnerable users. "Even when we sit down and go back and forth on like… okay, yes, in many contexts, this term is hate speech, it's more complicated than that for some groups," said P26. "And when our systems flatten a decision across all groups, we need to be really thoughtful about why—and what the consequences of that are."

### *1.5.2.3 Growth at all costs: Misalignment with business incentives*

Despite the challenges inherent to global scale, many contemporary businesses pursue what P5 described as a "growth at all costs" model, designed to increase profits and maximize shareholder value—creating strong incentives that frequently and directly conflict with typical Trust & Safety goals.

***Business incentives undermine safety goals.*** When asked about the most significant barriers to successful content moderation at scale, P11 said: "Shareholders, venture capital… I'm trying to stop short of saying 'capitalism.'" P5 agreed:

> "One way to look at the entire problem is as a negative externality of venture capital—and the venture capital model of funding platforms. You know, the 'growth at all costs' incentives that platforms have at various points in their lifecycles. I've talked with lots of students who have worked at various companies, and leaders at different platforms, and the folks I talk to always describe this process as an evolving nuisance to leadership. So I think there are some pretty big structural impediments to actually addressing this at scale."

These market pressures manifest in company-level goals, often requiring individual teams to prioritize user growth over user safety. "Company-level metrics like growth and engagement are in opposition with content moderation," said P3. "At Facebook, anything introduced to reduce harm also reduces engagement. You are running up against a top-line metric for the company that impacts the stock price—that everyone across the board is optimizing for."

Participants noted that content moderation efforts are often treated as compliance checkboxes, rather than sincere attempts to reduce harm. "A lot of the current goals of moderation are about compliance—making sure we don't have the bad content, because we'll get in trouble," said P4. "What would it look like instead for that to be about 'Are we actually hurting people?' I'm not saying platforms don't care about this at all, but a lot of things do come down to business incentives. If companies didn't have to make money, a lot of things would be really different." P3 agreed, drawing a parallel to for-profit healthcare:

> "It's very hard for us to incentivize ourselves to have the right outcomes in medicine if things are for profit. I think the same is true for social media companies. In a world in which we are trying to profitize interactions between people through 'growth' and 'engagement,' we'll never be able to solve some of these problems. If you're expecting a pharmaceutical company to grow year-over-year, that means more people need to be taking these drugs—which means you're solving less of the underlying problems, and just hoping to continue treating the symptoms, more and more every year."

***Metrics obsession drives optimization of numbers, not outcomes.*** While measurement is essential to understanding performance at scale, an overreliance on quantification and optimization can distort incentives, undermining broader goals. "When I'm trying to implement safety best practices, I often get pushback like, 'We basically only care about numbers,'" said P31. "They're very honest about that. Balancing different goals from people who are not safety-focused... it's so hard sometimes. It's really difficult."

Participants stressed the difficulty of developing meaningful metrics to assess complex and highly contextual harms—further widening the divide between Trust & Safety and business goals. "It's easier to report total takedowns, rather than—for example—total problems prevented by proactive interventions," said P13. P26 agreed:

> "Basically no one feels like they have good KPIs or measurements for success in this space. We're all really struggling… especially convincing the rest of your org that you are doing well. It doesn't mean that we're not trying, but it's an ongoing struggle."

Even where reliable, meaningful measurement is possible, user safety metrics are often in conflict with broader organizational goals. "It seems impossible to make the right decisions from a harm reduction standpoint if your goal is engagement, because harmful content is engaging," said P3. "It catches your eye—you want to read it; you want to see what's going on. A lot of times, that makes it really hard for these companies to actually take the hard step to reduce harm." Participants expressed similar concerns about common operational metrics, such as *average handle time* (i.e., the average duration of a content moderation decision), which encourage content moderators to prioritize speed over quality. "Productivity and accuracy metrics are one way of measuring, but it's probably not the best way," said P12. "That setup can lead to different types of shortcuts that eventually cause flaws in the system." P14 agreed:

> "They're under high productivity pressure—and what they do is game the system. They find ways to cut corners and make do with those productivity targets, which then leads to mistakes."

***Flawed "innovations" prioritized over fundamental safety.*** Participants felt that the rapid pace of product development within the technology sector prevents adequate risk mitigation, with innovation taking precedence over necessary product maintenance—including solutions to existing problems. P3 reflected on the difficulty of prioritizing foundational work in environments that incentivize novelty:

> "At Facebook, you're pretty much in a grind on the performance review cycle. A lot of times, the sort of like, baseline work isn't valued as much as flashy projects that get a lot of attention from a leadership level—so people drop a lot of the really hard but essential work to do content moderation well at scale. Things like working with the outsourcing sites. It's grueling work. It's very time-consuming. It requires a lot of time and energy.

> But that's not the work that's valued when you go through a performance review cycle—what's valued is a new feature launch or a large engineering project; something new and flashy. No one wants to do that work, and the people who do generally aren't rewarded for it in the right way."

Participants expressed particular frustration when companies shift focus to new product areas before fully addressing known problems in existing products. "I think another big issue is companies prioritizing or adding new areas of focus," said P24. "For example, how Facebook became Meta. There's this new huge focus on VR. It's still kind of crazy to me that there is this huge shift in focus when they haven't even solved the problems on the regular platforms." "There are constantly new products that we need to be supporting," added P8.

Participants emphasized the importance of comprehensive testing prior to a product's deployment to large populations of users. P8 expressed concerns about the rapid development of machine learning models, which he felt are not well understood even by those who build and maintain them:

> "One of the other challenges is, as we start using a lot more machine learning models, no one really understands how these models interact at scale. I keep hearing people say 'you need to release the algorithm.' There is no algorithm—there are hundreds of algorithms, and they all have emergent properties that we don't understand, even inside the company. Why was this decision taken? No one can explain that."

**Preoccupation with short-term costs prevents long-term gains.** Participants expressed frustration about the relentless pursuit of maximized earnings, particularly given regular spending reductions in areas deemed non-essential by company executives. "I don't really have an answer to the capitalism thing," said P19. "They're trying to run a large social media platform. There's always people who work in operations trying to make sure they're cutting costs in ways that can bring revenue into the company." Industry participants described company leaders as reluctant to allocate resources to initiatives that impact potential profits less directly. P28 said:

> "There's so much pushback, because it's like, 'But that costs money to change right now. I could spend the money later.' Yeah, but you could also know a thing is coming. If you know a thing is going to be better long-term, there needs to be better incentives to doing that thing, like harm reduction in moderation. If you can find a way to show where this harm reduction will, you know, not necessarily create better profits, but create better user experience—so that you have better user retention, so that we have better turnover rates—that's important to show, because then you can incentivize these things."

While the costs associated with scaled content moderation are immediate and easily observable, the benefits are both less explicit and cumulative, occurring over time—and only when moderation is successful. Moderation mistakes were characterized by participants as both inevitable and costly. "When you have something that has been moderated but maybe shouldn't have been—or something that maybe should have been moderated, but wasn't—the news of that can travel pretty quickly," said P19. "There are cascading effects of decisions that are done poorly, or just by the basic error of whoever's making the decision, that make it challenging to maintain trust in the governance of the platform." P27 agreed, emphasizing the fundamental imbalance between the longer-term shifts in community norms successful moderation can enable and the immediate costs of moderation mistakes:

> "There's an asymmetry inherent in scaled content moderation: if you moderate badly, the observed downsides are immediately apparent, in that people will see incorrect decisions being made. It leads to lack of trust in a pretty short-term time cycle—versus failing to moderate at all, the community can kind of maintain itself in the short term. The downsides in being hands-off only start to become apparent longer-term, and are much harder to quantify. That asymmetry leads to an implicit bias, where we tend to do nothing because it's easier to justify, at least in the short term. I think that's a very difficult and profound obstacle to overcome. There's always this subtle nudge to doing less than doing more, because we are more easily able to quantify the downsides of doing more."

***Constant deprioritization results in chronic underinvestment.*** This misalignment with foundational business incentives results in what industry participants described as chronic under-investment in content moderation and other work related to user safety. When asked about what could be done to improve the current state of scaled content moderation, P24 said: "Just investing more in integrity at these companies. This is a high-level value to users. We already are investing, but definitely not enough." P31 agreed: "Not enough money or employees focused on content moderation, combined with greed—tech companies and their motivation for existing, and motivation for scaling, how they prioritize growing as a company and where they dedicate their internal resources." P16 described challenges with even "basic resources, like having enough time to make evaluations."

Other participants questioned whether additional resources could sufficiently propel improvement without other, more systemic changes. "Something Amy Bruckman says in her book is that if we had unlimited resources, content moderation would be perfect—and I'm not sure I completely agree with that," said P4. "Some of the challenges here could not be solved by resources; things like values conflicts. In theory, a platform could hire enough people to look at every single piece of content within seconds that it's posted, and also pay them well and train them well and that sort of thing. Of course, that would be a massive amount of resources—and again, it wouldn't solve all of the problems." P10 agreed: "Facebook is a counterexample. Facebook has spent… it's a crazy amount of money. It's in the billions. They have tens of thousands of people; they've put in a lot of resources. But I don't know if the problem has gotten any better on Facebook. Some might say it's gotten worse." "And we're also constantly still hearing about poor conditions for the labor force," added P4. "So there could be more resources going into that, whether it's paying people more or mental health support."

Ultimately, participants overwhelmingly agreed that technology companies could and should be investing more in content moderation and other efforts to protect the safety of their users, highlighting the significant costs of underinvestment. P31 said:

> "If a company is not dedicating enough resources to cover all of its content moderation needs, the result of that, potentially, is massive widespread harm to society. We've seen how angry society is these days. Incels, violent extremism, all sorts of stuff online is causing real-world harm. Facebook has even come out with data that shows they knew they were damaging the mental health of teenage women on their platform. It's just really upsetting to think about how many companies are kind of intentionally sacrificing the well-being of society, for the sake of… not even making money, in some cases. Just numbers. We know a lot of these companies aren't making money—we're just collecting data. And at what cost?"

*1.5.2.4 Labor pains: Exploitation, isolation, and burnout*

Finally, participants reflected on challenges related to worker experiences, including exploitative labor arrangements, organizational siloing, and chronic burnout.

***Overreliance on third parties and exploitative working arrangements.*** Despite limited resources, Trust & Safety workers and related teams are expected to support continual platform expansion and scaled growth, resulting in what participants described as an overreliance on third-party labor. Technology companies have increasingly outsourced content moderation and other data labeling tasks to third-party vendor workforces, often in regions with lower labor costs (Gray & Suri, 2019; Roberts, 2019). Despite strict production quotas and regular exposure to disturbing content, individual contractors earn relatively low wages, without the benefits and job security associated with full-time employment. "Yeah, people are getting paid… for doing really, really hard work," said P2. "They're not getting paid enough. They're not being offered the benefits that they should be offered to do this very, very difficult work." P3 agreed, noting the disparity in working conditions between content moderators and other technology workers:

> "For content moderators, value them like you do an engineer, in terms of compensation and benefits. Treat them as human equals, as opposed to disposable people that you expect to get burned out from seeing harmful content. Value mental health and resiliency for these people that are actually having to view the content every day."

Participants expressed concern with the sustainability of outsourcing work that is critical to everyday business operations, in part because countries with lower labor costs are often located in regions more vulnerable to climate-related disasters. "Last year, there was a hurricane in the Philippines," said P23. "There was no power at all, throughout the whole country. There were no reviewers who could do anything. That is a very real problem, and irresponsible to ignore."

Companies may also expect that certain individuals will contribute their services on a voluntary basis. Civil society participants expressed particular frustration at the amount of work required to engage with platforms on behalf of users, which P22 described as "the free labor that people like me end up doing for tech companies." P22 continued: "As Trusted Partners[8], we have to mediate and escalate things that we see on the platforms, or they won't be taken down—really flagrant issues, like death threats, rape threats, homophobic content in different languages. We are supposed to do the content moderation work and escalation work to our public policy or human rights contacts at these companies, or it won't be taken down." In addition to advocating for users and flagging specific content for potential removal, civil society groups may also provide more foundational governance labor, often without compensation. P22, who compiled dictionaries of potentially harmful terms using regional expertise, expressed doubts about whether or not this work was ultimately implemented:

> "As someone from the Middle East, I have had to build lexicons for homophobic and transphobic language. We had to bring in groups from different countries, to build these lexicons for Facebook—but then we never see them applied. They say, 'Okay, we'll take this stuff and incorporate it in our language models.' I understand it takes time, but we do this work for free. Again, I just need to reiterate: for free. Because they don't want to invest—or they aren't necessarily as serious about investing—in language models and context-driven policies related to hate speech. We end up doing all the work and becoming traumatized."

---

[8] Meta maintains a network of Trusted Partners, comprising "over 400 non-governmental organizations, humanitarian agencies, human rights defenders and researchers from 113 countries around the globe" (Meta Transparency Center, 2023).

*Fractured understanding: Navigating organizational silos.* Outsourcing critical work also creates extreme siloing, where fundamental functions related to content moderation and broader platform governance are separated from their closest partners. When policy teams are organizationally distant from their operational counterparts, it results in what P23 described as "the gap between policy and protocol," which becomes particularly pronounced at scale. P23 continued:

> "Policy basically represents intent—what some platform considers harmful, potentially in consultation with some external experts. But then you need to translate that into a series of steps for the scaled review team. You can't just tell them, 'Hey, go nuts with your intuition,' because there would be no consistency whatsoever. That series of steps can have large or small gaps with the intent of the policy."

Participants also highlighted the impact of distancing full-time employees from their more precarious peers, including third-party content moderators and other contingent workers.

"Contractors are partitioned off somewhere else, where the rest of the company doesn't have good visibility and empathy on what they're doing," said P27. Industry participants described the daily experiences of content moderators as inaccessible to other employees—and, as a result, largely unaccounted for in the development of moderation policies, processes, and products, despite moderators possessing the most direct and timely knowledge of user dynamics and developing trends. "Content moderators don't feel they have a voice," said P22. "Even though they interact with this content the most and notice all the trends." For non-staff content moderators—for example, those users who voluntarily moderate a subreddit or Facebook Group—this gap is even more significant. "Community admins don't work with the Trust & Safety team at all," P6 said of Reddit. "So there's like, the people on the ground, working with the actual moderators, and then a whole other team working on what they call 'Anti-Evil Operations,' which is the system they use to implement site-wide community standards."

For content moderation to truly succeed at scale, participants expressed a need for what P14 described as "a more holistic system." P12 agreed, emphasizing the distance between user-facing product teams and the communities their products ultimately serve: "In the context of building a new tool, that is typically done by a team of engineers, led by a product manager—with very little input from the people who do the moderation work." While organizational distance between relevant teams is one key barrier, systemic understanding is also restricted by the fragmentation of outsourced labor and the precarity of contingent employment. P12 continued:

> "Even when such input is solicited, there are other kinds of organizational barriers to getting good information—including the fact that when you do a site visit to a moderation outsource vendor, you get the feeling that employees don't really feel free to express themselves fully, maybe under pressure from management. Of course, a bigger problem is that because of the atomized nature of the work they do, they might not even understand the types of questions that are being posed to them by the people who want to build tools, so there are several barriers."

***Burnout, turnover, and institutional knowledge loss.*** Left with inadequate resources and facing relentless growth, individual workers—whose efforts to ensure user safety at scale present challenges under even ideal conditions—are left feeling unimportant, overburdened, and isolated from their closest peers, resulting in what participants describe as chronic burnout and frequent employee turnover. "There's a very high burnout and turnover rate for content moderators in general, but also for the people who work on these topics at the company," said P24. "At Facebook, a lot of people don't last more than a year on some of these teams, or leave the company altogether."

While the mental health and overall wellness of front-line moderators has garnered increasing attention in recent years (e.g., Roberts, 2019), participants expressed similar concerns for Trust & Safety workers and others in comparable roles. Although these workers are typically (though not always) full-time employees—and as such enjoy significantly more institutional support than their outsourced peers—they, too, conduct challenging work with insufficient resources. In addition to performing their role-specific responsibilities, workers may face challenges to their emotional and psychological well-being, such as what P6 described as "emotionally managing how you feel after dealing with difficult content." Participants indicated that Trust & Safety workers may also face unique administrative barriers, with some organizations requiring additional approval processes before certain employees can proceed with potentially sensitive work. "You're getting pushback from Legal on research or projects that you want to launch that could have huge impact," said P24.

These disproportionate impacts contribute to difficulties recruiting and retaining talent. "It does take time to build out teams—and to keep people on teams," said P24. "Burnout prevention for FTEs and content moderators… we know this is very difficult stuff to work on." P2 highlighted the difficulty of retaining expertise when individual workers regularly move between organizations:

> "It didn't matter how many people platforms were hiring, especially at the major companies. There just never seem to be enough people hired. I've heard more recently at the majors that there's a huge retention problem as well. This is a new field. These companies, as they're building up, they're trying to get talent—and so they're pulling from all these other companies. It's creating a pipeline problem. There just doesn't seem to be enough people, in terms of creating policy but also enforcing it. It doesn't matter how many tens of thousands of people are hired; it never seems to be enough."

In addition to problems retaining talent and appropriately resourcing critical teams, companies experiencing rapid growth may favor hiring processes that predominantly emphasize process efficiency, at the potential expense of identifying highly skilled candidates with domain-specific expertise. "When companies scale, they inevitably sacrifice on hiring requirements and training," said P31. "Part of the problem is, yes, it's a young industry. I worry over time, when things balloon, you have less focus on training—less focus on hiring people who can think through things in a nuanced manner."

Ultimately, participants emphasized that technology companies do not appropriately value work relating to user safety. Several participants drew parallels to other care-based industries, such as healthcare and domestic labor, where essential work is similarly undervalued. P16 expressed "wanting to see care work valued more highly; things like the 'Wages for Housework' campaign. Trying to push for the valuing of what would otherwise be under-resourced labor." P12 agreed, lamenting what she described as a general "undervaluing of the importance of the work—and workers." "Sometimes, some of the employees—and I think especially leadership—tend to forget that these are actual people doing the content moderation," said P24.

***Executives enjoy disproportionate influence.*** While individual workers suffer the implications of increasingly precarious work, company executives enjoy disproportionate institutional power—including significant influence over individual content moderation decisions. Participants suggested that certain company leaders may overestimate the relevance of their own experiences and instincts, ultimately influencing organizational direction and broader strategy. "I remember very clearly when Dick[9] had gotten feedback from all of his friends that Block wasn't a feature that Twitter needed," said P9. "They really just needed Mute, so they should roll out Mute and get rid of Block at the same time. I was like, 'That is the dumbest idea, and this will end in tears for many people.' It only lasted like six hours after rollout, but it was very illustrative."

When creating visual representations of current scaled content moderation processes, P8 and P9 highlighted the influence of certain "priority input," including what they described as "random issues encountered by friends of the C-suite." Other participants reflected on the influence of governments and other political actors on executive decision-making. P3 described ethical challenges faced by workers when asked to act in ways that conflict with their team's stated mission and goals:

> "I think this is also something that is hard on ICs. I worked on escalations for a couple of years when I first started. A lot of times, we'd get escalations from foreign governments asking us to take down 'fake accounts' that were critical of the government. There were numerous examples of situations where leadership—I'll be intentionally vague here—would say, like, 'Do this, because we need to appease the government.' And everyone who is an IC working on this was like, 'What do we do? This goes against our team's premise. This isn't a fake account.' We're being asked to take it down as a fake account, and it's really only because it's an account that's critical of the government. And depending on the government, the company would push back or not push back. Morally, everyone always felt very conflicted about it."

---

[9] Dick Costolo is a former Twitter COO (2009–2010) and CEO (2010–2015).

Participants emphasized a disconnect between the teams responsible for identifying risks to user safety and those with sufficient authority to make strategic decisions. "Ultimately, the communication flows one way—it comes top-down," said P14. "It's super hard for engineering or policy to convince CEOs of anything, based on any kind of bottom-up understanding of a problem." In the absence of meaningful input, individual workers are often left to implement or even justify decisions they had no power to influence.

### 1.5.3 Content moderation futures: What can be done?

After identifying numerous ways in which contemporary technology companies fail to successfully govern their platforms, participants considered potential solutions.

#### *1.5.3.1 Embrace holistic visions*

In order to achieve sustainable success, participants encouraged technology companies to adopt a more holistic vision for the role of content moderation in broader community health. Participants highlighted that contemporary moderation systems are largely reactive, with the majority of governance-related resources allocated to the adjudication of individual content decisions—a phenomenon P8 and P9 described as "the 'whack-a-mole' element." P10 agreed:

"Moderation systems are really just thinking about individual content decisions, and not like, what are the other things we're trying to promote? What does promoting health actually look like? I think that that is a big blocker, at least in the companies I've seen. They are pretty overwhelmed with having to make that decision, you know, hundreds of thousands of times, over and over and over again, in this kind of never-ending queue."

Instead of investing in reactive strategies and short-term fixes, participants emphasized the need for company leaders to clearly articulate a holistic vision of product safety and health. "There's just constant fracturing in the moderation space," said P33. "There's fracturing within a single area; there's fracturing between areas. There really does need to be a top-down vision that pulls it together. Leadership has to have a long-term, 'This is the thing we're trying to build' view—and it needs to be one that doesn't just rely on a single solution. Even though my colleagues have attempted to offer that repeatedly, I have yet to see someone in leadership adopt a vision like that at a social media company."

*Moderation is essential.* Participants believe the success of scaled content moderation relies on a paradigm shift: one that recognizes moderation as an integral component of the product experience, rather than a crisis response or otherwise necessary nuisance. Industry participants observed resistance to the characterization of content moderation infrastructure as foundational platform architecture. "Leadership is a little bit adversarial," said P33. "They view moderation as something that needs to be done—as a last resort, rather than an essential." P12 agreed: "I would reorient leadership to understand that good moderation is, in fact, a value add—it's not simply a burdensome 'cost center' to a company. It could be reconfigured and thought about in a better way—to actually be a selling point for a platform, rather than something to be hidden and swept under the rug."

Elevating the position of governance labor also requires improving the working conditions of individual employees. "If I could wave a magic wand and just change people's mentality, I would do things like remove stigma and organizational barriers," said P12. "The first thing I would probably eliminate is the system of outsourcing this labor in the first place, and pay it an appropriate wage for the skill level that it requires, which is very high." P14 agreed, highlighting the ways in which low wages and limited career advancement further marginalize content moderation work at the organizational level:

> "The standing of content moderation within companies… it's an afterthought. It's like, we have growth, we have engineers, we have sales, and then we have additional 'also needed, but not important'-type things—like catering, housekeeping, and content moderation. Like, someone has to deal with the complaints, right? There isn't really a vision of the importance of community, or content moderation's place within the actual product—or a vision of a social network at all, considering that these are the biggest platforms for humans to communicate. That doesn't seem to have resonated with any of the leaders of these organizations. And that's also seen in the career path of the content moderation guys. You have these super difficult and nuanced policies that are to be applied, ideally with 100% accuracy, by some outsourced people who get low wages. It's debatable if that's ever going to work."

This structural devaluation is intensified by broader labor and equity concerns present across industries. "There are some macro-economic changes that need to be made," said P2. "What we're basically trying to do is figure out how we can create an equitable system, where maybe we have fewer of these problems. People are feeling a little safer, we have fewer divisions, so there are incentives for equitable and ethical moderation. Things like higher wages, the 3- or 4-day work week, paid leave… other ways of introducing equity in society. Universal healthcare. Broader socioeconomic changes would be helpful—not just reproducing inequitable conditions."

*Moderation is forever.* Participants emphasized that content moderation, like other forms of governance, is a perpetual necessity—it can never be permanently solved. "There's a sort of, 'Oh, we have this problem solved, so we can stop worrying about it,'" P8 said. "Not realizing that it's an adversarial relationship. You never have that problem solved." P33 agreed:

> "Leadership is thinking that eventually they can stop—like if we invest enough, we'll have some magical solution, and then we can walk away from it. But moderation is forever fundamental to the product. It will have to be maintained forever. There's this weird magical thinking about it that I wish would go away."

Industry participants felt that company leaders largely desire 'silver bullet' solutions, even for complex sociotechnical problems that lack a simple, scalable fix. "There's a lot of like, 'Everything is on fire,'" said P11. "Leadership wants a single solution—once you have it, it's done. It's essential to move toward Trust & Safety people engaging with content moderation concerns as part of their product development. Moving toward things not just being on fire, but where employees are in a sustainable place; where they're actively and persistently building toward preventing harm, with organizational and financial support." P33 agreed, advocating for more integrated approaches to the design and management of safety-related products and interventions:

> "There are logistical burdens that show up in the health space that don't get addressed—like the fact that companies have more than one product surface they're trying to solve for. They let the surfaces solve their problems separately, so there's often a lot of noise and siloing. They're designing one-off strategies—like 'Hey, we'll remove the content'—when no single strategy is going to be effective by itself. It needs to be a set of them, but people struggle to think beyond the one strategy on their mind. The way products are getting built puts a lot of additional burdens on the people actually attempting to build these systems. A lot of the leadership involved really fails to provide—or even allow the development of—holistic missions that would help tie together different parts of products and unify how they're working."

***Moderation is communal.*** Participants described moderation as a critical mechanism for cultivating community, which requires consideration for a variety of individual experiences. "Content moderation has to be about the safety and the health of the greater community," said P7. "You should care about every individual who is involved." Promoting health at the community level requires consideration for the broader impacts of an individual's actions, which may not always align with their intentions. "For successful moderation at scale, it doesn't matter how the content conceivably could be understood," said P1. "What matters is how people are actually understanding it." P7 agreed:

> "You see people who are like, 'I didn't mean it that way.' I get it, but I cannot care what your intent was if your impact was that everybody had a bad day. That's a problem. If I'm moderating 1,000 active chatters, I cannot care if one person is having a bad day. I have to respect the fact that there are 999 other people who would like to have a good day."

Participants also discredited the notion that content removal is tantamount to censorship, emphasizing the role of moderation in maintaining an environment conducive to safe expression. "The inappropriate positioning of content moderation as an issue of 'free speech' is derailing and inaccurate," said P12. "People in positions of power conflate 'free speech' with a right to be abusive, violent, and toxic towards others," added P21. P33 compared content moderation to other necessary maintenance of public spaces:

> "Moderation promotes free speech. I brought up the public park metaphor: you want people to use your public park? Okay, well, people don't want to walk on the grass if there's glass in it—so you limit activities that will make glass in it. And that actually makes more people want to go there and do more things."

*1.5.3.2 Design platforms that encourage prosocial behavior*

Participants stressed the role of platform design in shaping user behavior, particularly for preventing or lessening harm. Because current approaches to scaled enforcement fall short of stated goals, participants described alternative approaches influenced by other governance contexts, such as criminal justice reform.

***Establish clear expectations for appropriate behavior.*** Instead of focusing the majority of governance resources on policing rule-breakers, participants encouraged platforms to first establish clear expectations for appropriate behavior. "How do we shape community behaviors and norms?" asked P17. "There's a lot of focus on rule-breaking and the consequences of that. We have less on the norm-shaping side of it." While certain users may knowingly violate platform rules, some misbehavior can be prevented by ensuring that users understand normative expectations. P4 said:

> "Do research on what the rules should be, how they should be presented to people, and how people learn them. What would a moderation system look like if the biggest goal was to help people learn the rules? That is clearly not the intended goal of the majority of moderation systems right now, particularly at scale; the goal is for there to not be bad content on the platform. Helping people learn the rules could be a way to accomplish that, but it's seen as a more roundabout way. The typical way to accomplish that is to identify and delete that content—and if people happen to learn the rules along the way, okay."

Users are regularly exposed to normative signals that influence their understanding of a platform and shape their future usage, including their interactions with other users. In the absence of explicit guidance about platform rules and expectations, direct observations of others' behavior will be the predominant normative influence. P10 emphasized the importance of onboarding experiences in setting clear expectations for new users, who are often hastily ushered into creating and engaging with content:

> "It starts when someone joins a platform or some new space. It's hard to talk about the stuff that happens at the very end—the moderation—without thinking about how they got there. You join a space, and you are shown some set of rules. Sometimes that's like, 'Hey, go read our rules,' or sometimes, 'Here's our Terms of Service; click Accept,' but rarely is it nuanced onboarding, where someone is guided through what the norms and expectations are. And pretty quickly thereafter, you're encouraged to create some content, whether that's filling out your profile or creating a post. Equally quickly, you're exposed to a bunch of content—'Hey, here's a bunch of people you should be friends with or connect with.' A lot of it is just observing what other people are doing, because people often don't jump in and just start creating a bunch of stuff. They're often just lurking, at least for the beginning, and starting to see how other people are engaging in this space."

This brief exposure to platform rules also typically occurs before new users have enough knowledge of the platform to appropriately contextualize them. "The first thing that happens related to moderation is that, as soon as you join, you are—in theory—shown the rules," said P4. "You don't learn the rules. You're shown the rules… maybe. It's so divorced from when you actually create content. It's not like when you post your first piece of content, then you're told about the rules." Because new users also lack an established network and personalized feed, their first experiences of a platform will largely be characterized by algorithmic recommendations of content that is engaging, but not necessarily normatively appropriate or even typical. P10 said:

> "What if there was more curation around people's first exposure to the space? Most of the time, you're just thrown in—you're exposed to whatever you're exposed to. And on platforms that have algorithmically filtered feeds, you're being shown content that tends toward, like, salacious or controversial. The closer things get to pushing up against the rules, the more engagement and eyeballs it tends to get—so that tends to be what people are seeing. That doesn't seem to be conducive toward helping people learn what the rules are."

***Promote rehabilitation, not retribution.*** Because expectations for appropriate user behavior are often unclear, participants felt that platforms should prioritize user education over potential penalties. "I would love for content moderation systems to incorporate a 'user education first' approach," said P25. "With the exception of high-severity harms or malicious intent, the system should not default to enforcement first, but to training or education." P6 agreed, highlighting the lack of opportunities for offending users to understand and correct their behavior: "Part of what's challenging is that there seem to be so few opportunities for people to learn from their mistakes. For people that do end up violating a policy or norm, it can be really challenging."

Many contemporary platforms also lack specific incentives to encourage or reward desirable behavior, further contributing to uncertainty about normative expectations. "There's a lack of incentives for users to behave well," said P20. "Especially in commercial content moderation, which often uses punishment. On Reddit, they have awards and karma to incentivize users to behave well." Given that users lack definitive guidance about appropriate conduct, participants felt that certain violators deserve more leniency. P25 advocated for a "second chances approach" for first-time violators, who rarely reoffend:

> "Many users who violate do so once—and after some education, do not continue to violate. This suggests that there should be the opportunity for users to 'undo' the problematic behavior, before any enforcement or strikes would kick in. If this hasn't occurred within, like, 24 hours, the platform steps in and a strike is applied."

P10, who recommended "more research about the experience of people who break rules," agreed that most policy violators are not repeat offenders—and discouraged platforms from ascribing malicious intentions to every user who runs afoul of their policies. "I'm constantly in these rooms where people are talking about 'bad actors,'" said P10. "I've done multiple recidivism analyses across different platforms that show it's an exceedingly small percent of people, despite how much room in the conversation they take up."

***Favor flexible interventions.*** Not all rule violations are equal, and participants also suggested that platforms should implement more contextual interventions that better account for differences between violation types and offender motivations. "It has to do with context," said P18. "How do we understand a piece of content without also looking at the user, and what it might mean for that particular user in their own context?" P6 advocated for penalties that are proportionate to the specific offense, including the violation history of the offender:

> "Depending on what the violation is, context is taken into consideration. Different outcomes depend on what the violation actually was, and what kinds of information we know about the user. Somebody who's repeatedly violating a rule is probably not going to be dealt with the same way as somebody who's done it for the first time, or somebody who is brand new to the community. The severity of the infraction, the potential for harm that it causes… all of that is going to depend."

Uniform penalties also fail to account for differences between harmful behaviors, with acute behaviors often treated identically to those that progressively worsen. "We know that there are some problem areas—like eating disorders, suicide, self-injury, radicalization and others—that can have an escalation trajectory," said P25. "There is a percentage of the population where their behavior will become increasingly worse over time, but the response by platforms tends to be the same each time there is a harmful event." P8 agreed: "Harm is usually a process, not an event—and we only look at single events."

Participants also felt that many platforms rely on overly blunt interventions, which P9 described as "a historical over-indexing on content removal as being the sole type of content moderation." "Moderation is often focused on very binary output," agreed P30. "Leave up and remove." By focusing on binary removal decisions, platforms may lose sight of broader governance goals, such as promoting civil discourse or reducing user harm. P10 said:

> "There is an orientation toward this binary of moderation, where stuff is either good or bad. Especially in the actual operations of content moderation, binary decisions are being made: stuff stays up, or it gets taken down. I think it points us toward the wrong goals.

There are many other goals a moderation system can work toward. Trying to address harm is something that doesn't really serve, when you're just thinking about whether something stays up or goes down."

Ultimately, participants stressed that effective interventions encourage offenders to reform their behavior. "Effective sanctions actually correct behavior that is untoward," said P27. "We struggled at Facebook coming up with ideas that actually move the needle toward creating a healthy community and aren't just overly simplistic or heavy-handed… that end up doing more damage than good."

*Implement responsive regulation.* P33 shared a pyramid model influenced by Braithwaite's (2002) responsive regulation (see Figure 1.2), which allows for the deliberate application of adaptable sanctions with escalating penalties to encourage cooperation. "This is another way of thinking about different types of content—a holistic vision," said P33. "Not just using a single strategy to moderate content, but using a collection of strategies, and gradually layering on more strategies as risk of harm becomes more severe. It's very focused on a more restorative justice view." Because responsive enforcement systems provide offenders with opportunities to adjust their future behavior, users who repeatedly violate rules are clearly signaling their intention—creating justification for the application of harsher penalties, such as device-level bans. Instead, most contemporary platforms favor blunt enforcement systems, applying uniform penalties to every offender.

Participants noted that uniform penalties are both too harsh to effectively reform well-intentioned offenders and too lenient to deter chronic recidivists. "There's a lack of ability in some cases to tell whether you're dealing with somebody who really just needs rehabilitation and is sort of… fixable, let's say," said P29. "Or if you're dealing with somebody who is there to game your systems, and it's like the seventh account that they've created; they're just maliciously wasting your time and engaging in bad faith. When you have 'one size fits all' systems, you end up struggling to design systems in a way that optimizes for both use cases— and you end up pleasing neither." P32, who encouraged platforms to "crack down harder on repeat offenders," suggested that platforms who fail to appropriately penalize frequent violators may facilitate their continued influence:

"We've been working on this project about what happened to the election lie superspreaders—who were the biggest spreaders of election fraud in 2020, and what ever happened to them online? What we learned is only a third of them were banned. The rest have gone on to be completely divisive in a series of issues that have divided the country since. They went on to be influential in conversations about grooming and critical race theory—and yet, there was an opportunity where they really did break the rules in a grave way. If there were a harder crackdown on those particular megaphones, there would be a different conversation."

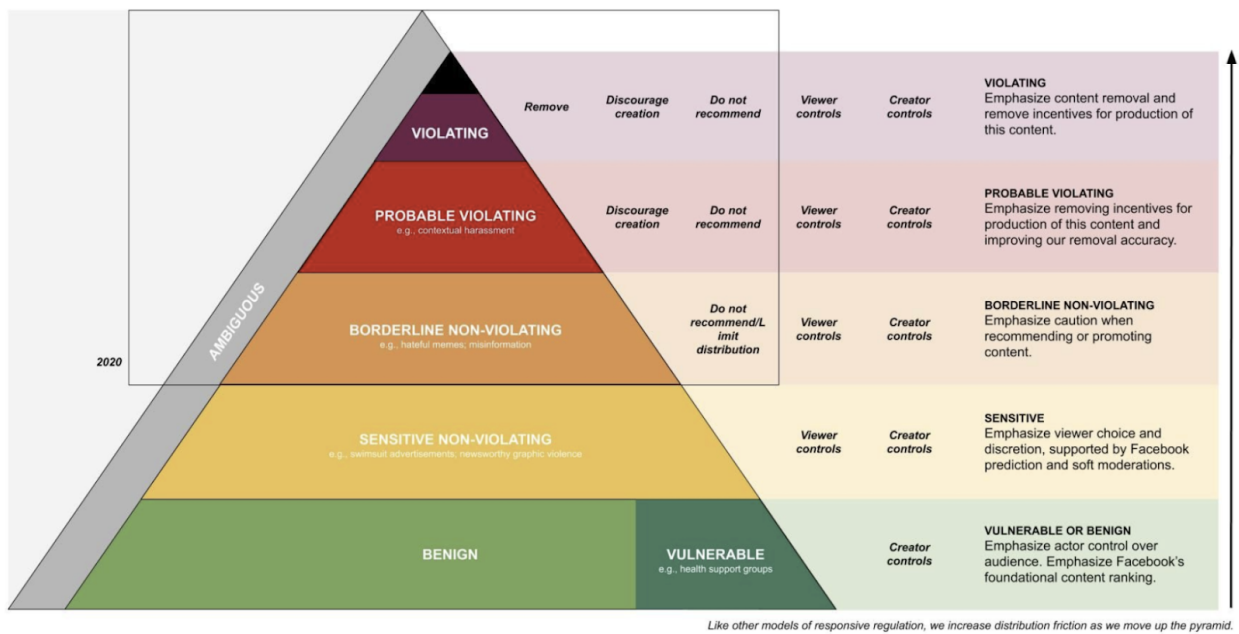

Figure 1.2: An enforcement model influenced by Braithwaite's (2002) responsive regulation.

### 1.5.3.3 Incentivize corporate accountability

Existing corporate structures have failed to adequately protect users' safety. Participants stressed the need for greater corporate accountability, whether incentivized via regulatory structures or alternative levers. "Companies are not accountable," said P20. P32 agreed:

"I do think there is a bigger role for governments to play. Private companies are in the business of making decisions they probably shouldn't be making on their own. Too much of the public square is being outsourced to these private companies. It's become a job—but these are major societal issues that historically governments would play in much more."

***Regulation: Promising, but not a panacea.*** Participants described regulation as a critical mechanism for encouraging corporate responsibility, including basic transparency into company practices. "When I first started reporting on this in 2017, Facebook wouldn't even tell us that they had content moderators," said P32. "It was so unglamorous, it was like they wouldn't even acknowledge it—until the Congressional hearings." P31 noted the effectiveness of financial penalties in motivating compliance with child safety laws and other public interest policies:

> "Capitalism could potentially remove incentives for tech companies to invest in certain areas. If you've seen Silicon Valley, there's a famous episode where one of the characters almost bankrupts the company by messing up COPPA, the Children's Online Privacy Protection Act. Basically, there are massive fines from the government if you are found to not be deleting people under the age of 13 from the platform, or if you are negligently allowing them onto your platform in a way where you're collecting their data. I would love for there to be more laws that motivate companies to remove more misinformation, to be more proactive… to maybe even discourage companies from existing, if they cannot exist in such a way where they're not causing societal harm. I'm in favor of government intervention, while being mindful of the risks associated with its misuse."

While participants expressed support for the development of new regulations to force more responsible corporate action, they questioned lawmakers' proficiency in a problem space rife with complexity and nuance. "Emerging regulation is becoming really important at the moment," said P25. "Not all of the specific regulations are actually fully informed on the realities and challenges of content moderation." This disconnect between policymakers and technologists may result in clumsy or ineffective regulations that fail to address the root cause of these problems—while simultaneously creating additional stress for resource-sparse teams. P9 referenced Australia's Criminal Code Amendment (Sharing of Abhorrent Violent Material) Act 2019, which passed through both houses of parliament in less than 48 hours (Douek, 2020):

> "I can tell you, from massive amounts of personal experience, the challenges involved in getting even a group of governments in a set of countries that are generally aligned to agree on anything within a time span of less than, I don't know, a two-year period. It's like herding cats on methamphetamines. And the cats are going to also just leave and then do something like what Australia did—they wrote a law and then passed it overnight, basically, in the aftermath of the Christchurch shooting."

Participants also expressed discomfort with the potential for government overreach and corruption, particularly given the surveillance value of user data. "Do we want to be outsourcing this to governments?" asked P32. "Governments around the world have deep interests in targeting citizens for certain reasons. Many governments shouldn't be trusted with that." P9 replied:

> "I think that's a robust understatement. If you look at the rankings for press freedoms, even the countries where theoretically they're quite high, you still have almost inevitably bad actors within those systems who could still be trying to push for removal of content that is completely legitimate—whistleblowers, activists, dissidents. We've had examples of governments straight up telling us untruths to try to get content removed, like claiming there wasn't a protest taking place. There absolutely were protests happening. The government was fully aware of them, and they were just trying to shut that down before it got out further."

P31 agreed, acknowledging the political and moral complexity of government intervention, which may effectively reduce certain harms while also creating additional risks:

> "It's very complicated. Should we potentially have laws or even higher-level governmental systems in place that restrict access to certain content? I hate that idea. I read 1984. But at the same time, I look at countries that do have some restrictions in place, and their CSAM risk is almost eliminated in some cases—so I do think that some government intervention might be necessary. I worry that tech companies are sacrificing the well-being of society for greed; for making money. I want more government intervention and more laws, without those laws or tools being exploited by corrupt government officials. I don't know what the solution is."

While external incentives may be necessary to reorient broader corporate priorities, internal change is still needed. Regulation is a powerful lever for motivating company leaders to dedicate resources to areas of their businesses they may otherwise neglect—but as P7 noted, systemic transformation requires cultivating additional understanding and commitment:

> "We talked about finding ways to incentivize companies to prioritize users—and user safety—over corporate profits. Some of this can be government pushes; laws and guidelines, like 'If you don't do this, we will take your profits' or whatever. If you incentivize companies to do that, you can then take the time to do things like educate—everybody at a company, not just moderators, but all the way up to your C-suite. Why is content moderation important? Why is nuance important? You can have incentives to make companies not care as much about their profits, which is great, but if there is no training—if there's no push for nuance—then you lose any kind of forward momentum that you get from having laws that will inflict punishments on folks."

Participants also cautioned that technology companies may be incentivized to identify loopholes to avoid increased scrutiny by governments and other regulatory agencies. "I truly believe that Facebook chose to go to end-to-end encryption to avoid regulation," said P21. "It had very little to do with protecting people who are vulnerable; there are plenty of other tools that journalists, activists, and other people in marginalized situations can use. They wanted end-to-end encryption for Messenger to avoid responsibility."

***Leverage alternative accountability mechanisms.*** Creating true accountability for modern technology companies will require leveraging a variety of mechanisms. In addition to pursuing regulatory solutions, participants emphasized the importance of public scrutiny, including negative media attention. "The incentive for firms to do good content moderation is really a soft power incentive," said P12. "Sure, there are government regulations—and there are more and more coming online—but outside of the cases that those regulations address, which are a fairly small percentage, what really ends up being the incentive for firms is things like bad PR, or a risk of losing their userbase." P2 agreed, noting that profit can be an effective lever for motivating corporate responsibility:

> "Commercialization adds some good things towards this dynamic. Some of the ways we've really gotten movement on a lot of moderation issues is because of like, trying to keep advertisers happy. That can be both bad and good; it just depends on which groups you're making happy, and those commercial logics. The real incentive is to keep people on the platform."

Pressure from users can be an effective lever, too. "Reddit describes it as a process of maturing," P2 continued. "Actually trying to figure out what these dynamics should be, and drawing some clear lines around subreddits like r/fatpeoplehate or r/beatingwomen. Really what that was was a lot of pressure from the community. Pressure from advertisers and things like that." Still, participants recognized the stark power differentials between even organized communities of users and other corporate stakeholders, such as investors. "There's nothing really binding platforms to be responsive to the communities that get harmed by them," said P5. "Nothing like the way they have to be responsive to their investors and SEC filings. There's nothing that's really tightly coupling those groups." P12 agreed:

> "In terms of users, the leverage is just not there in an organized way—in the way it is there for shareholders. The case study of this power imbalance is what has gone on with Twitter, because nothing about that was good for users. Nothing about that was good for the public. And yet fiduciary duty to shareholders was what guided that entire process, and what will continue to guide it. We see a real misalignment there."

Ultimately, corporations wield power that individual users lack, and participants emphasized the importance of structured coordination and collective action in provoking meaningful change. Persuading technology companies to be more accountable to their users may require creative solutions for consolidating individual influence, such as what P5 described as an "eyeball strike":

> "Imagine all the users of Facebook somehow unionize and drive revenue to zero. I was thinking of ways to upend the system. If you have no accountability to the communities that you harm, unions are a historical example of an approach to mitigate that."

*1.5.3.4 Explore alternative models*

Corporate accountability is not guaranteed, and participants also emphasized the need to explore alternative technology models that allow for greater transparency, experimentation, and user empowerment.

***Forsake the 'killer app.'*** Participants reflected on the fundamental tension between establishing consistent, principled governance and the pursuit of continual expansion and growth. "Platforms are trying to operate so broadly, in terms of who they want on there," said P2. "It makes it very difficult to create one rule. You could be alienating lots of different parts of your community."

When technology companies design platforms intended to serve heterogeneous populations of global users, certain values are necessarily prioritized over others, resulting in governance practices that inevitably marginalize some users. "Many of these platforms are designed to create one large public space," said P17. "And the consequence is that there are certain groups that are more privileged than others. It creates lots of weird disincentives for effective moderation." Rather than committing to specific principles, most platforms pursue impartiality, in the hopes of appealing to the broadest possible userbase. When they fail to establish universally applicable rules, they also compromise their ability to deliver consistent scaled outcomes. "There was a reluctance from platforms, understandably, to intervene in geopolitics," said P2. "So there was a problem with creating one kind of rule that could be applied more generally—which is a problem if you're trying to moderate at scale." P32 added:

> "I feel like it could be more efficient if each platform just said its values. That's actually one of the interesting things about some of the platforms that have emerged on the right. They say they're 'free speech' platforms, but in a way, what they are is a particular ideology—and they're okay with it. I know that would sacrifice users, if you didn't at least try to appeal to everybody… but there is no societal common denominator, and so admitting that, and then going 'Okay, well what's *our* common denominator, even though it's going to make a lot of people disagree with us.' Getting away from that approach of trying to be all things to all people would be helpful."

Participants also highlighted the difficulty of governing broad, general-purpose environments designed to serve a range of user needs. "We really run into issues with spaces designed for the flexibility to become many things," said P26. "Because we've instilled that flexibility, we need to collapse and standardize. In spaces that are designed for very particular outcomes, we can do a lot more to empower individual users or small groups to manage their own safety and to escalate requests for help—because those spaces don't suddenly change."

Instead, participants encouraged companies to design technologies for specific purposes and audiences. "One thing that I would love to see a lot more of—and I think empowers a lot of better solutions for addressing disruptive or harmful content—is having platforms and spaces designed so that they are better calibrated for the needs of that group," said P26. "Maybe the fundamental problem is just that we're using the wrong platforms," said P29. "Maybe the mistake we're making is trying to find a way to make one platform be everything for everyone."

*Empower community governance.* Given the challenges associated with uniform governance of general-purpose platforms, participants advocated for more localized and participatory alternatives, empowering users to shape and steward their own community spaces. "Top-down content moderation interventions can cause harm," said P11, recommending "more empowerment on the user side." Even companies that utilize scaled, platform-level governance rely heavily on volunteer moderation labor, such as users who voluntarily moderate Facebook groups or subreddits. Participants recognized the unique burdens on these users, who spend significant amounts of their personal time interpreting user reports, coordinating with other moderators, and communicating outcomes to offenders—often without the benefit of automated enforcement tools or other operational mechanisms common in scaled moderation. P19, who researches user-governed Discord servers, encouraged platforms to implement moderator rotation programs, both to reduce the possibility for moderator burnout and create additional opportunities for users to participate in community governance:

> "People want to create their own community management—but they do seem to get trapped sometimes. They can't give up the power; and then, of course, a lot of people get burnt out. Incentives for rotating people in and out helps with a lot of the problems of moderating, especially at scale, for these kinds of self-governing spaces. And not just to solve the problem of delinquent moderators or burnout, but also… what would it mean for the community if everyone has an opportunity to help out and make it a better place?"

Participants reflected on the value of shared responsibility, noting the importance of community engagement in promoting ownership and accountability among individual members—in turn reducing potential misbehavior. "If I can just go to a community garden and pick up trash, that's one way I could help contribute," said P19. "I'm thinking of online spaces in a similar way. It doesn't require monetary incentives. There is a stronger reward, in that I also get to have stronger participation." Other participants questioned who ultimately benefits from this vision of collective community stewardship, particularly under platform capitalism, where user-generated content is heavily monetized. "I really love that," said P2. "Except it would be so weird if Monsanto came into that community garden and started selling all of the vegetables the community was producing. That's kind of the dynamic right now."

## 1.6 Discussion

These results reveal that contemporary social media governance is broken—yet few companies seem to meaningfully invest in its repair. I argue that successful governance is undermined by the pursuit of technological novelty and rapid growth, resulting in platforms that necessarily prioritize innovation and expansion over public trust and safety. To counter this dynamic, I revisit the computational history of care work, to motivate present-day solidarity amongst platform governance workers and inspire systemic change.

### 1.6.1 Move slow and fix things

Modern technological development is shaped by the promise of "innovation," informed by the techno-optimistic notion of technological progress as both inevitable and desirable (Avle et al., 2020). Innovation is tightly coupled with entrepreneurialism, which encourages individuals to contribute to broader economic development by promoting a democratizing ethos of invention, collapsing "vast gaps in money, formal knowledge, and authority" (Irani, 2019) to disguise the power dynamics that determine who will succeed and who fails (Becerra & Thomas, 2023). Innovation promises novel solutions to human problems, but also the production of wealth for individual innovators, complicating distinctions between public good and private gain. Participants in this study observed how this ideology manifests within large social media companies, describing corporate cultures that prioritize new product development over the resolution of existing product harms—reflecting an incentive structure that rewards speed, novelty, and growth over user safety.

Innovation is frequently framed as a self-evident social good, emphasizing disruption for the sake of progress and disguising the political and economic interests embedded in what, how, and for whom innovation occurs (Irani, 2023). By casting technological advancement as universally beneficial, the potential harms produced by unbridled innovation—such as labor displacement, algorithmic discrimination, and expanded surveillance—are obscured. As a result, the prevailing ideology of innovation normalizes risk-taking while externalizing responsibility for risk mitigation, reinforcing a model of progress that privileges economic ambition over public accountability and societal health. Participants described how this logic legitimizes underinvestment in platform governance, resulting in Trust & Safety teams who lack sufficient resources or institutional authority to produce effective moderation. Instead of recognizing governance as a platform's core function (Gillespie, 2018), key responsibilities are delegated to underpaid commercial moderators, unsupported community volunteers, and uncompensated third-party experts—practices that reflect a broader strategy of outsourced accountability.

In order to industrialize inventive production, innovation encourages frequent experimentation, resulting in the rapid deployment of technology products with minimal testing and limited attention to potential downstream consequences (Pfotenhauer et al., 2021). Commercial content moderation systems rely heavily on machine learning models, which participants noted are frequently released without robust internal understanding of their actual behavior—which may differ from their intended function. Even those responsible for the development of automated tools are unable to explain how specific governance decisions are made by the systems they built and maintain. This orientation toward efficiency comes at significant cost: inaccurate moderation, inconsistent policy enforcement, and the marginalization of users whose contexts and experiences fall outside the narrow training data used to calibrate these automated systems.

Participants similarly emphasized a lack of corporate investment in foundational governance infrastructure. While "new and flashy" projects receive attention and accolades from company leaders, participants noted that "hard but essential work"—such as maintaining or improving content moderation tools—receives little recognition during evaluations of individual and team performance. This orientation toward innovation disincentivizes investment in content moderation, system maintenance, and other essential safety infrastructure, areas characterized by company leaders as cost centers rather than engines of growth. The result is a corporate culture that rewards the continual creation of new features, products, and platforms, producing "innovation" at a pace that restricts even internal scrutiny—and resulting in the public release of experimental technology products with unknown ethical implications.

Success in innovation is commonly defined by how quickly and widely a product can be deployed, reflecting an underlying alignment with market-driven values. By prioritizing speed and scalability, innovation favors actors with existing resources and infrastructure, constraining which problems are solved and for whose benefit (Irani, 2019; Irani, 2023). Problems that cannot be solved through scalable solutions—or whose solutions primarily produce social but not monetary benefit—are deemed less relevant or even antithetical to corporate goals. Participants noted the "adversarial" positioning of efforts to mitigate risks to user safety, due to perceived delays in product development timelines or negative impacts to key performance metrics. More holistic, forward-looking governance strategies and pro-social platform designs—"innovations" rooted in social justice and collective success—are repeatedly sidelined, despite both theoretical promise and empirical validation (Tyler et al., 2021; Katsaros et al., 2022). Taken together, these findings reveal how the prevailing ideology of innovation externalizes the labor of governance and delegitimizes repair, rendering the systemic work of harm reduction peripheral to the business of building scaled platforms.

**1.6.2 Countering the politics of scaling**

Scalable platform technologies embody a broader "scalability zeitgeist," or the modernist preoccupation with technological solutions that can be efficiently replicated at exponential scales (Pfotenhauer et al., 2021). Contemporary ambitions of scale are articulated through entrepreneurial strategies such as "growth hacking," or the rapid acquisition of users to accelerate revenue growth—privileging speed and market dominance over product quality and longer-term sustainability. Economies of scale have long been a cornerstone of industrialization, exemplified by the mass production of standardized goods, enabling high output at low cost. The rapid industrialization of commercial content moderation is no different, and as platforms expand, they rely on scalable, standardized workflows and low-cost labor to process vast volumes of user-generated content quickly and efficiently, often at the expense of equitable governance (Gillespie, 2018; Roberts, 2019).

While scalability is often marketed as a sustainability strategy—with the purported aim of allowing businesses to handle increasing demand without incurring proportionally higher costs—it signifies a particular vision of the future grounded in perpetual growth (Hardy, 2019), made possible by the externalization of social and environmental costs. In what Pfotenhauer et al. (2021) describe as the "Uberization of everything," countless technology start-ups now aspire to disrupt existing markets by building scalable platforms that shift costs, risks, and other traditional business responsibilities onto others—namely users, workers, and governments. As social media platforms similarly prioritize growth and scalability, participants described chronic underinvestment in Trust & Safety teams and related functions, increasingly shifting the burden of platform governance to outsourced workforces, algorithms, regulators, and users themselves.

Scalability creates structural constraints on what kinds of governance are possible. Due to the demands of global scale, content moderation systems become reactive by default, producing an unrelenting stream of individual decisions. This Sisyphean effort—which participants likened to an endless game of "whack-a-mole"—requires extensive resourcing, impeding investment in more preventative solutions. These politics of scaling profoundly shape the governance capacity of contemporary social media platforms, as evidenced by the findings presented in this paper, which underscore the structural tension between global scalability and the lack of universally applicable standards for acceptable behavior. While effective governance is responsive to context—for example, local variation in norms, values, and interpretations of harm—participants describe ways in which scale inherently obscures difference, resulting in content moderation policies and practices that cannot adequately address nuanced harms. Instead, social media platforms implement crude solutions that predominantly address issues with broad consensus, in what one participant described as a "lowest common denominator" approach to moderation—producing governance outcomes experienced by users as arbitrary, inconsistent, or unjust.

Under what Srnicek (2017) describes as platform capitalism, platform technologies generate profit by building and controlling digital infrastructures that mediate interactions between users—for example, connecting riders with nearby drivers—while continuously extracting, analyzing, and monetizing the data those users produce (Hardy, 2019). By positioning themselves as intermediaries, platforms avoid the legal and regulatory responsibilities of direct service providers (Gillespie, 2010) while accumulating massive amounts of valuable user data (Zuboff, 2018), which they use to optimize engagement, predict behavior, and deliver targeted advertising in increasingly automated ways. These logics of extraction result in platform designs that tolerate or even encourage harmful content or behavior in pursuit of profitability, which participants described as a fundamental misalignment between business incentives and user safety—exacerbated by a culture of quantitative obsession that cannot meaningfully capture or effectively respond to complex, contextual harms.

Though the imperative to scale is frequently presented as a neutral or purely technical goal, this framing obscures the social, economic, and political stakes of widespread expansion (Hanna & Park, 2020; Pfotenhauer et al., 2021). Scalable governance demands uniform rules, and participants emphasized the disproportionate influence of Western perspectives on content moderation policies and practices, reflecting computing's intrinsic "colonial impulse" (Dourish & Mainwaring, 2012) and further marginalizing vulnerable users by reproducing global power asymmetries and entrenching structural harms. Through the lens of scalability, social media platforms become sites of contested governance, where decisions about speech, visibility, and harm are shaped by capitalist logics of extraction and expansion. Without a fundamental reorientation toward equity, accountability, and care, the structures that enable corporate profit will continue to undermine the conditions necessary for just and effective governance.

### 1.6.3 Reclaiming computational logics of care

As the present study demonstrates, improving the future state of scaled content moderation will require fundamentally reorienting how governance is framed—not as a reactive system of control, but as a continuous, contextual, and communal practice of care. Though dominant narratives of technological progress often invoke masculine, capitalist logics of innovation, disruption, and expansion (Irani, 2015a; Irani, 2019), technology is created and sustained by feminized and therefore politically devalued forms of care labor, such as support, maintenance, and repair (Fisher & Tronto, 1990; Jackson, 2014). Indeed, computing itself was built by care labor: NASA's core memory was constructed by "little old ladies" who hand-threaded thousands of wires through small magnetic cores, delicate craftwork requiring precision and patience—yet rendered "unworthy of remembrance" by the male engineers and astronauts who both depended on and were credited with its success (Rosner et al., 2018).

As Rosner et al. (2018) argue, the characterization of this highly skilled, error-intolerant work as manual ("feminine, menial, low-status") rather than cognitive ("masculine, innovative, high-status") labor demonstrates a lack of sympathetic context, or the situated knowledge required to appropriately value sociopolitically invisible labor (Star & Strauss, 1999). In the present work, participant experiences reveal that corporate technology leaders similarly fail to recognize the complexity of governance work, consistently undervaluing the labor required to responsibly govern online platforms. While participants characterized content moderation as an ongoing obligation, they questioned "magical thinking" from company executives, who expect scalable, permanent solutions—and who demonstrate a general reluctance to invest in any governance labor. In what one participant described as an "Oh Shit" model of governance, companies typically only establish a content moderation program following a crisis, instead of designing platforms with prevention and sustainability in mind.

Governance labor is also rendered structurally invisible by resisting the forms of measurement that typically justify institutional investment—and when value cannot be easily counted, it is systematically discounted. Because Trust & Safety work is ill-suited to quantification, corporate practices that prioritize the optimization of standardized metrics—used to communicate quarterly performance to company shareholders, but necessarily abstracting "the process of work being done" (Star & Strauss, 1999)—result in what participants described as chronic underinvestment in essential safety infrastructure. Despite content moderation tools that participants emphasized lack necessary context, the maintenance and improvement of existing systems is not prioritized, producing moderation outcomes that are at best inaccurate, and at worst, inequitable. Initiatives that do not readily translate into measurable business outcomes— such as increased user engagement or advertising revenue—are similarly deprioritized, reflecting a narrow conception of value that ignores the long-term benefits of community-driven design. These results reveal how foundational caretaking labor is rendered legible only as cost, not value, reflecting a broader political economy that separates socially reproductive labor—the labor of maintaining people, communities, and institutions—from labor considered economically "productive" under capitalism (Fraser, 2016).

Under financialized capitalism, short-term financial gain is privileged over long-term production, including social well-being—despite capitalist production relying on broader social capacities (Fraser, 2016). Social platform technologies exhibit a similar contradiction: in place of the slow, situated, and relational labor required to sustain healthy individuals and communities, technology companies privilege binary enforcement mechanisms that can be automated and quantified, mistaking efficiency and throughput for attentiveness and care. Content moderation professionals described current scaled moderation systems as overly reactive and reliant on blunt interventions (e.g., "leave up or remove"), despite workers' intimate knowledge of—and desire to produce—more intentionally prosocial designs. As Seering et al. (2022) argue, content removal is just one element of "a deeper social process of nurturing, overseeing, intervening, fighting, managing, governing, enduring, and stewarding communities," a perspective reflected in the range of care-based interventions—user education, rehabilitation, responsive regulation, and so on—recommended by participants. Rather than merely policing infractions to achieve minimal compliance with regulatory requirements, participants expect technology companies to take seriously the work of cultivating, repairing, and sustaining the communities their products purport to serve.

This refusal to appropriately value care work is compounded by structural power imbalances, including labor precarity. Much like the women who built NASA's core memory systems, commercial content moderators are viewed as expendable, even as they absorb the psychological and operational burdens of the platform's most difficult problems (Roberts, 2016; Roberts, 2019). Participants recounted the precarious conditions of outsourced workers, who earn low wages, lack mental health support, and work remotely from regions with increasingly volatile climates. While contractors are physically and socially distanced from the designers and engineers who build the platforms they maintain, civil society partners voluntarily contribute expertise they rarely see integrated into public-facing products. Even salaried Trust & Safety workers conduct challenging work while organizationally isolated from their closest partners, resulting in what participants characterized as chronic burnout and frequent employee turnover. Platform governance workers operate in silos, with little influence over the systems they are charged with protecting—a structural fragmentation that reinforces the invisibility and undervaluation of care labor.

These findings reveal a platform governance ecosystem that depends on care labor but refuses to honor it, with complex, skillful, and emotionally demanding work obscured behind dashboards, outsourced across borders, or mistaken for a compliance task rather than a core function. Content moderation, like all forms of governance, is a manifestation of care, and it deserves to be valued accordingly. Reclaiming computational logics of care requires foregrounding the experience of workers (Irani, 2015b; Roberts, 2019), investing in maintenance and repair (Jackson, 2014; Schoenebeck & Blackwell, 2021), and building systems that reflect the relational nature of governance infrastructure (Star & Ruhleder, 1996; Gillespie, 2018). As feminist HCI scholars have long argued (Bardzell, 2010; Irani, 2015b; Dombrowski et al., 2016; Rosner et al., 2018), to care is to attend to complexity, and to commit to the critical evaluation of technologies out of an interest in their material improvement—or as Puig de La Bellacasa (2011) writes, "we must take care of things in order to remain responsible for their becomings."

One potential antidote to this ongoing crisis of care (Fraser, 2016) is the organization of social media governance labor, resisting institutional atomization and centering the workplace as an essential location for contesting the possibilities of governance. Aligned with recent calls for the promotion of a worker-centered HCI (Irani, 2015b; Dombrowski et al., 2017; Rosner et al., 2018; Roberts, 2019; Fox et al, 2020), these results underscore the importance of scholarly attention to the governance workers who build, maintain, and protect the platforms that structure our daily lives—and the need for individual workers to organize their collective power. Organized care labor is visible care labor, a critical first step in transforming scaled content moderation from a series of fragmented tasks into a shared practice with a common voice, collective memory, and strategic leverage. As HCI expands its ethical commitments beyond usability and inclusion, centering the working conditions and organizing capacity of governance workers across the "sociotechnical stack" (Qiwei et al., 2024) offers a pathway toward redistributing power and normalizing the value of care. Ultimately, while improving the future state of scaled content moderation will require a host of systemic changes, worker solidarity is foundational to achieving equitable governance and broader social justice.

## 1.7 Conclusion

This study broadens current understandings of social media governance by examining the lived experiences of practitioners who enact, examine, and engage with scaled content moderation systems. Through a series of participatory design workshops with content moderation professionals—from part-time content moderators and university researchers to corporate vice presidents—I produce a more intimate understanding of the complex landscape of people, practices, and politics that ultimately determines how contemporary social media platforms are governed. By characterizing platform governance as essential care labor—highly skilled, constant, and undervalued—this research highlights the need for worker solidarity and broader structural change to advance more equitable technology futures.

### 1.7.1 Acknowledgements

Thank you to Umang Bhojani for his contributions to data collection and workshop design. I am deeply grateful to my research participants for their competence, candor, and courage. I also wish to acknowledge those who felt unable to safely participate, and whose absence is not overlooked.

Dombrowski, L., Alvarado Garcia, A., & Despard, J. (2017). Low-Wage Precarious Workers' Sociotechnical Practices Working Towards Addressing Wage Theft. In *Proceedings of the 2017 CHI Conference on Human Factors in Computing Systems*, 4585–4598.

Domínguez Hernández, A., Ramokapane, K. M., Das Chowdhury, P., Michalec, O., Johnstone, E., Godwin, E., . . . Rashid, A. (2023). Co-Creating a Transdisciplinary Map of Technology-Mediated Harms, Risks and Vulnerabilities: Challenges, Ambivalences and Opportunities. *Proceedings of the ACM on Human-Computer Interaction, 7*(CSCW).

Douek, E. (2020). Australia's "Abhorrent Violent Material" Law: Shouting "Nerd Harder" and Drowning Out Speech. *Australian Law Journal, 94*(1), 41–60.

Dourish, P., & Bell, G. (2011). *Divining a Digital Future: Mess and Mythology in Ubiquitous Computing.* The MIT Press.

Dourish, P., & Mainwaring, S. D. (2012). Ubicomp's Colonial Impulse. In *Proceedings of the 2012 ACM Conference on Ubiquitous Computing*, 133–142.

Ehrenreich, B., & Ehrenreich, J. (1977). The Professional-Managerial Class. *Radical America,* 11(2), 7–32.

Fisher, B. & Tronto, J. C. (1990). Toward a Feminist Theory of Caring. In E. K. Abel & M. K. Nelson (Eds.), *Circles of Care: Work and Identity in Women's Lives*, 36–54. State University of New York Press.

Foucault, M., & Nazzaro, A. M. (1972). History, Discourse and Discontinuity. *Salmagundi, 20*, 225–248.

Fox, S. E., Khovanskaya, V., Crivellaro, C., Salehi, N., Dombrowski, L., Kulkarni, C., . . . Forlizzi, J. (2020). Worker-Centered Design: Expanding HCI Methods for Supporting Labor. In *Extended Abstracts of the 2020 CHI Conference on Human Factors in Computing Systems*.

Fraser, N. (2016). Contradictions of Capital and Care. *New Left Review, 100*(99).

Gillespie, T. (2010). The Politics of 'Platforms'. *New Media & Society, 12*(3), 347–364.

Gillespie, T. (2018). *Custodians of the Internet: Platforms, Content Moderation, and the Hidden Decisions That Shape Social Media.* Yale University Press.

Goldman, E. (2020). *A Pre-History of the Trust & Safety Professional Association (TSPA).* Technology & Marketing Law Blog. http://blog.ericgoldman.org/archives/2020/06/a-pre-history-of-the-trust-safety-professional-association-tspa.htm
69

Irani, L. (2015b). Hackathons and the Making of Entrepreneurial Citizenship. *Science, Technology, & Human Values, 40*(5), 799–824.

Irani, L. (2019). *Chasing Innovation: Making Entrepreneurial Citizens in Modern India.* Princeton University Press.

Irani, L. (2023). Encountering Innovation, Countering Innovation. *Engaging Science, Technology, and Society, 9*(2), 118–130.

Irani, L. C., & Silberman, M. S. (2013). Turkopticon: Interrupting Worker Invisibility in Amazon Mechanical Turk. In *Proceedings of the SIGCHI Conference on Human Factors in Computing Systems*, 611–620.

Jackson, S. J. (2014). Rethinking Repair. In T. Gillespie, P. J. Boczkowski, & K. A. Foot (Eds.), *Media Technologies: Essays on Communication, Materiality, and Society,* 221–239. The MIT Press.

Jiang, J. A., Middler, S., Brubaker, J. R., & Fiesler, C. (2020). Characterizing Community Guidelines on Social Media Platforms. In *Companion Publication of the 2020 Conference on Computer Supported Cooperative Work and Social Computing*, 287–291.

Jungk, R., & Müllert, N. (1987). *Future Workshops: How to Create Desirable Futures.* Institute for Social Inventions.

Katsaros, M., Tyler, T., Kim, J., & Meares, T. (2022). Procedural Justice and Self Governance on Twitter: Unpacking the Experience of Rule Breaking on Twitter. *Journal of Online Trust and Safety, 1*(3).

Kawakita, H., Nishimura, M., Satoh, Y., & Shibata, N. (1967). Neurological Aspects of Behçet's Disease: A Case Report and Clinico-Pathological Review of the Literature in Japan. *Journal of the Neurological Sciences, 5*(3), 417–439.

Keller, D. (2022). *The EU's new Digital Services Act and the Rest of the World.* Verfassungsblog.

Keller, D. (2023). Platform Transparency and the First Amendment. *Journal of Free Speech Law, 4*.

Keller, D., & Leerssen, P. (2020). Facts and Where to Find Them: Empirical Research on Internet Platforms and Content Moderation. In N. Persily & J. Tucke (Eds.), *Social Media and Democracy: The State of the Field and Prospects for Reform,* 220–251. Cambridge University Press.
71